\documentclass[12pt]{iopart}

\newcommand{\citethesisboeltzig}{\cite{thesis:boeltzig_blinded}}

\renewcommand{\citethesisboeltzig}{\cite{thesis:boeltzig_unblinded}}

    \usepackage{graphicx} 
    \usepackage{caption}
    \usepackage{hyperref} 
    
    \usepackage[separate-uncertainty=true, range-phrase={ -- }]{siunitx} 
    
    \newcommand{\nuclide}[2]{\ensuremath{{}^{#1}\mathrm{#2}}}
    \newcommand{\plusp}{\ensuremath{+ \mathrm{p}}}
    \newcommand{\pg}{\ensuremath{(\mathrm{p},\gamma)}}
    \newcommand{\pa}{\ensuremath{(\mathrm{p},\alpha)}}
    \newcommand{\ag}{\ensuremath{(\alpha,\gamma)}}
    \newcommand{\an}{\ensuremath{(\alpha,\mathrm{n})}}
    
    \newcommand{\reaction}[5]{\nuclide{#1}{#2}#3\nuclide{#4}{#5}}
    \newcommand{\shortreaction}[3]{\nuclide{#1}{#2}#3}    

    \newcommand{\Ex}{\ensuremath{E_\mathrm{x}}} 
    \newcommand{\Ep}{\ensuremath{E_\mathrm{p}}} 
    \newcommand{\Ecm}{\ensuremath{E_\mathrm{c.\,m.}}} 
    
    \newcommand{\Erescm}{\ensuremath{E^{res.}_\mathrm{c.\,m.}}} 
    
    \newcommand{\Thalf}{\ensuremath{T_{1/2}}} 
    
    \newcommand{\eg}{e.\,g.\@}
    \newcommand{\ie}{i.\,e.\@}
    
    \newcommand{\insitu}{in-situ}
    \newcommand{\exsitu}{ex-situ}
    
    \newcommand{\LUNAfourhundred}{LUNA-400}
    \newcommand{\TantalumOxide}{\ensuremath{\mathrm{Ta}_2\mathrm{O}_5}}
    \newcommand{\betaplus}{\ensuremath{\beta^{+}}}
    \newcommand{\Qvalue}{$Q$-value}
    \newcommand{\gammaray}{$\gamma$-ray}

\begin{document}

    \title
    [Advances in Radiative Capture Studies at LUNA]
    {Advances in Radiative Capture Studies at LUNA with a Segmented BGO Detector}


    \author{
        J.~Skowronski$^{1,2}$,
        R.\,M.~Gesu\`e$^{3,4}$,
        A.~Boeltzig$^{5,6}$,
        G.\,F.~Ciani$^{7,8}$,
        D.~Piatti$^{1,2}$,
        D.~Rapagnani$^{3,4}$,
        M.~Aliotta$^{9}$,
        C.~Ananna$^{3,4}$,
        F.~Barile$^{7,8}$,
        D.~Bemmerer$^{5}$, 
        A.~Best$^{3,4}$,
        C.~Broggini$^{2}$,
        C.\,G.~Bruno$^{9}$,
        A.~Caciolli$^{1,2}$,
        M.~Campostrini$^{10}$,
        F.~Cavanna$^{11}$,
        P.~Colombetti$^{12, 11}$,
        A.~Compagnucci$^{6,13}$,
        P.~Corvisiero$^{14,15}$,
        L.~Csedreki$^{16}$, 
        T.~Davinson$^{9}$,
        R.~Depalo$^{17,18}$,
        A.~Di\,Leva$^{3,4}$,
        Z.~Elekes$^{16}$,
        F.~Ferraro$^{17,18}$,
        A.~Formicola$^{19}$,
        Zs.~F\"ul\"op$^{16}$,
        G.~Gervino$^{12, 11}$,
        A.~Guglielmetti$^{17,18}$,
        C.~Gustavino$^{19}$,
        Gy.~Gy\"urky$^{16}$,
        G.~Imbriani$^{3,4}$,
        M.~Junker$^{6,13}$,
        M.~Lugaro$^{20,21}$
        P.~Marigo$^{1,2}$,
        E.~Masha$^{5,17}$,
        R.~Menegazzo$^{2}$,
        V.~Paticchio$^{8}$,
        R.~Perrino$^{8}$\footnote{Permanent address: INFN Sezione di Lecce, Lecce, Italy},
        P.~Prati$^{14,15}$,
        V.~Rigato$^{10}$,
        L.~Schiavulli$^{7,8}$,
        R.\,S.~Sidhu$^{9,22,23}$,
        O.~Straniero$^{24,19}$,
        T.~Sz\"ucs$^{16}$,
        S.~Zavatarelli$^{15,14}$
        }

    \address{$^1$ {Dipartimento di Fisica}, {Università degli Studi di Padova}, {{Via F. Marzolo 8},  {35131} {Padova}, {Italy}}}

    \address{$^2$ INFN, {Sezione di Padova}, {{Via F. Marzolo 8},  {35131} {Padova}, {Italy}}}

    \address{$^3$ {Dipartimento di Fisica ``E. Pancini''}, {Universit\`a degli Studi di Napoli ``Federico II''},  {{Complesso Universitario di Monte Sant'Angelo - Via Cinthia, 21}, {80125} {Naples}, {Italy}}}

    \address{$^4$ INFN, Sezione di Napoli, {{Complesso Universitario di Monte Sant'Angelo - Via Cinthia, 21}, {80125} {Naples}, {Italy}}}

    \address{$^5$ {Helmholtz-Zentrum Dresden-Rossendorf}, {{Bautzner Landstra\ss{}e 400}, {01328} {Dresden}, {Germany}}}

    \address{$^6$ INFN, {Laboratori Nazionali del Gran Sasso}, {{Via G. Acitelli 22}, {67100} {Assergi}, {Italy}}}

    \address{$^7$ {Dipartimento di Fisica ``M. Merlin'', {Università degli Studi di Bari ``A. Moro''}, {{Via E. Orabona 4}, {70125} {Bari}, {Italy}}}}

    \address{$^8$ INFN, {Sezione di Bari}, {{Via E. Orabona 4}, {70125} {Bari}, {Italy}}}
    
    \address{$^9$ SUPA, Scottish Universities Physics Alliance, School of Physics and Astronomy, University of Edinburgh, EH9 3FD Edinburgh, United Kingdom}
    \address{$^{10}$ Laboratori Nazionali di Legnaro, Viale dell’Università 2, 35020, Legnaro (PD), Italy}
     
    \address{$^{11}$ INFN, Sezione di Torino, Via P. Giuria 1, 10125 Torino, Italy}
    
    \address{$^{12}$ Dipartimento di Fisica, Universit\`a degli Studi di Torino, Via P. Giuria 1, 10125 Torino, Italy}
    
    \address{$^{13}$ Gran Sasso Science Institute, Viale F. Crispi 7, 67100 L'Aquila, Italy}

    \address{$^{14}$ Università degli Studi di Genova, Via Dodecaneso 33, 16146 Genova, Italy}
    
    \address{$^{15}$ INFN, Sezione di Genova, Via Dodecaneso 33, 16146 Genova, Italy}
    
    \address{$^{16}$ Institute for Nuclear Research (ATOMKI), PO Box 51, H-4001 Debrecen, Hungary}
    
    \address{$^{17}$ Università degli Studi di Milano, Via G. Celoria 16, 20133 Milano, Italy}

    \address{$^{18}$ INFN, Sezione di Milano, Via G. Celoria 16, 20133 Milano, Italy}
    
    \address{$^{19}$ INFN, Sezione di Roma, Piazzale A. Moro 2, 00185 Roma, Italy}
    
    \address{$^{20}$ Konkoly Observatory, Research Centre for Astronomy and Earth Sciences (CSFK), E\"otv\"os Lor\'and Research Network (ELKH), Konkoly Thege Mikl\'os \'ut 15-17, H-1121 Budapest, Hungary; MTA Centre of Excellence}
    
    \address{$^{21}$ ELTE E\"otv\"os Lor\'and University, Institute of Physics, Budapest 1117, P\'azm\'any P\'eter s\'et\'any 1/A, Hungary}

    \address{$^{22}$ GSI Helmholtzzentrum f\"ur Schwerionenforschung, Planckstra\ss{}e 1, 64291 Darmstadt, Germany}
    
    \address{$^{23}$ Max-Planck-Institut f\"ur Kernphysik, Saupfercheckweg 1, 69117 Heidelberg, Germany}
 
    \address{$^{24}$ INAF-Osservatorio Astronomico d'Abruzzo, Via Mentore Maggini, 64100, Teramo, Italy

}
    
    \ead{a.boeltzig@hzdr.de}
{}


\vspace{10pt}



\begin{abstract}
    Studies of charged-particle reactions for low-energy nuclear astrophysics require high sensitivity, which can be achieved by means of detection setups with high efficiency and low backgrounds, to obtain precise measurements in the energy region of interest for stellar scenarios. High-efficiency total absorption spectroscopy is an established and powerful tool for studying radiative capture reactions, particularly if combined with the cosmic background reduction by several orders of magnitude obtained at the Laboratory for Underground Nuclear Astrophysics (LUNA).
    We present recent improvements in the detection setup with the Bismuth Germanium Oxide (BGO) detector at LUNA, aiming to reduce high-energy backgrounds and to increase the summing detection efficiency. The new design results in enhanced sensitivity of the BGO setup, as we demonstrate and discuss in the context of the first direct measurement of the \SI{65}{keV} resonance ($\Ex = \SI{5672}{keV}$) of the \reaction{17}{O}{\pg}{18}{F} reaction. Moreover, we show two applications of the BGO detector, which exploit its segmentation. In case of complex \gammaray{} cascades, \eg{} the de-excitation of $\Ex = \SI{5672}{keV}$ in \nuclide{18}{F}, the BGO segmentation allows to identify and suppress the beam-induced background signals that mimic the sum peak of interest. We demonstrate another new application for such a detector in form of \insitu{} activation measurements of a reaction with \betaplus{} unstable product nuclei, \eg{}, the \reaction{14}{N}{\pg}{15}{O} reaction.
\end{abstract}

    \vspace{2pc}
    \noindent{\it Keywords}: \gammaray{} total absorption spectroscopy, nuclear astrophysics, segmented BGO detector, background reduction, passive shielding, efficiency enhancement, activation technique
    
    \submitto{\jpg}
    
    \maketitle


\section{Introduction}
\label{sec1}

The small cross sections and weak resonance strengths that govern the astrophysical reaction rates in stellar environments translate to low experimental yields in the laboratory. Direct cross section measurements therefore require high beam intensities and large target densities to increase the reaction yield, in combination with highly sensitive detection setups. Achieving a high sensitivity for radiative capture reactions requires both a high detection efficiency for the signature \gammaray{}s of the reaction and a low rate of the background events that mimic precisely this signature. The detection efficiency depends on the detector type and size as well as the detector-target geometry and materials in between. The impact of the backgrounds can be reduced by lowering the rate of background events, as well as by exploiting experimental signatures that are more specific to the reaction of interest and less susceptible to other sources.

Deep underground laboratories provide unique conditions for key experiments in nuclear astrophysics, thanks to their dramatic reduction of the background originating from cosmic radiation. Depending on the energy region of interest, the cosmic background can be reduced by many orders of magnitude, resulting in drastically enhanced sensitivities \cite{aliotta_exploring_2022}. The combination of a deep underground location and a high intensity accelerator has been the foundation of long successful campaigns at LUNA \cite{Broggini-2018}, and motivated the construction of several new deep-underground accelerator facilities -- CASPAR \cite{Dombos_2022}, JUNA \cite{JUNA}, LUNA-MV \cite{LUNA-MV} -- as well as a shallow-underground accelerator laboratory, the Felsenkeller \cite{Felsenkeller}.
With the cosmic background greatly suppressed, other background sources take center stage. Typical candidates are environmental or intrinsic radioactivity, and beam-induced reactions on contaminants in the target. Further background reduction therefore requires targeted experimental efforts. Added shielding and the selection of radiopure materials are examples of changes in the experimental setup, whereas beam-induced backgrounds can be reduced by improved chemical purity, or discriminated against with the help of certain detector designs.

In this article we focus on recent improvements of a Bismuth Germanium Oxide (BGO) detector setup for radiative capture studies at LUNA, based on changes in setup and data analysis, with the goal to further enhance its sensitivity. We present how two reaction studies motivated these upgrades, and which opportunities come with the improved performance.
An introduction to the BGO solid target setup and the Monte Carlo simulation tools used throughout this work is given in subsections \ref{sec:1.1} and \ref{sec:1.2}, respectively.
Section \ref{sec:2} details the recent improvements in target and shielding setup at LUNA, and the resulting enhancement in efficiency and background suppression.
Section \ref{sec:3} is focused on the performance of the BGO setup in view of the first direct measurement of the elusive \Erescm{ = \SI{65}{keV}} resonance in the \reaction{17}{O}{\pg}{18}{F} reaction \cite{Buckner-2015}.
An innovative application of the BGO detector to measure \betaplus{} decays is reported in Section \ref{sec:4}.
The conclusion and an outlook on future applications of this setup are given in Section \ref{sec:5}.

\subsection{Total Absorption Spectroscopy with a BGO Detector at LUNA}
\label{sec:1.1}

Installing the \SI{400}{kV} accelerator \cite{Formicola-2003} at LUNA opened the path to experiments that provided a wealth of low-energy nuclear data for astrophysics. Total Absorption Spectroscopy (TAS) experiments have been a cornerstone of radiative capture reaction studies at LUNA, employing a highly efficient Bismuth Germanium Oxide (BGO) detector \cite{casella_new_2002} with extremely low background levels for reactions with high \Qvalue{}s. BGO was chosen for its large average atomic number and density, which allows for the high \gammaray{} detection efficiency in a relatively compact setup, crucial for measurements requiring massive shielding.
Additionally, two TAS detectors have taken up operation recently in other deep-underground laboratories: a new BGO-based detector has recently been commissioned and used for first first measurements at JUNA \cite{zhang_direct_2021, Zhang_19F_pg_2022}, and the HECTOR detector made of NaI(Tl) was transported underground for measurements at CASPAR \cite{HECTOR-CASPAR, Dombos_2022, Shahina_direct_2022}.

The first description of the LUNA BGO detector is given in \cite{casella_new_2002}, in the context of the measurement of the \reaction{2}{H}{\pg}{3}{He} reaction cross section, which covered the energies of interest for the Sun (solar Gamow peak) for the first time \cite{casella_old_2002}.
In brief, the LUNA BGO detector consists of six optically independent segments, each of them read out by one photomultiplier tube. By adding the energy of coincident events in the individual crystals, a total energy (sum, or add-back) spectrum can be obtained, while the energy deposition in the individual crystals allows to infer information on the individual \gammaray{}s emitted in the cascades. 
In the now more than 20 years of its operation at LUNA, the BGO detector was utilized in experiments employing a variety of different shielding-detector-target combinations, motivated by a range of science cases,  see Tab.\ref{tab:Tab1-summary_setups}. Often times, outstanding sensitivity was the requirement for these measurements, such as for the first direct measurement of the \reaction{14}{N}{\pg}{15}{O} cross section at stellar energies delivered with unprecedented accuracy on a windowless gas target \cite{lemut_first_2006, bemmerer_low_2006}. Another milestone was the first direct observation of the \SI{92}{keV} resonance in \reaction{25}{Mg}{\pg}{26}{Al} on a solid target, with a reported resonance strength as low as $\omega\gamma = \SI{2.9(6)e-10}{eV}$ \cite{strieder_25mgp_2012}. The latter resonance has recently been studied at JUNA \cite{su_first_2021}.
More recently at LUNA, the BGO detector in combination with a neutron shielding was used for a first direct study of the tentative \Erescm{= \SI{334}{keV}} resonance in \reaction{22}{Ne}{\ag}{26}{Mg}, with a sensitivity that allowed to establish an upper limit for its $\omega\gamma$ of \SI{4.0e-11}{eV} \cite{piatti_first_2022}.
Setup, data acquisition and analysis techniques evolved continuously throughout the decades with this detector, further pushing the limits of experimental sensitivity.  The work presented here is a continuation of the efforts reported in \cite{boeltzig_improved_2018}, and guided by the background models presented in that work. Of the two beam lines at \LUNAfourhundred{} we will here focus on the solid target setup.

In parallel to reducing the background rates in the detector,
suitable analysis techniques of the data from the detector are crucial to reach the highest possible experimental sensitivity. Whilst the outstanding sensitivity achieved in early LUNA experiments, with the BGO detector simply in summing mode, the detector segmentation does provide the means for a more refined data analysis, depending on the scientific case of interest.
As the six detector crystals are optically independent, the individual signal acquisition chain allows to record each event with timestamp and energy information. In the offline analysis the add-back spectrum is reconstructed with a coincidence window of \SI{3.5}{\mu\second} (cf.~\cite{boeltzig_improved_2018}). The advantages of TAS were emphasized above. However, a clear disadvantage is that in summing mode, the BGO detector does not yield direct information on the \gammaray{} cascades contributing to the sum peak (such as their branching ratios, \gammaray{} energies, or multiplicities). By exploiting the segmentation of the detector, one can infer additional information on the individual \gammaray{} energies. For example, the signature of the \reaction{22}{Ne}{\pg}{23}{Na} low-energy resonances decay scheme was, indeed, recovered by gating on the ROI, corresponding to the excitation energy of interest, in the add-back spectrum \cite{ferraro_high-efficiency_2018}. The single crystal spectrum obtained in this way was then fitted in order to determine the branching ratios.
Thanks to the large efficiency of the BGO setup, it was possible to determine the contribution of weak $\gamma{}$-transitions, that were not observed in the earlier LUNA campaign with HPGe detectors \cite{Cavanna-2015}. 
Gating the add-back spectrum can also significantly reduce the beam-induced background, which does not share the same de-excitation cascade as the reaction under investigation.
It is evident that exploiting the BGO segmentation offers a powerful analysis tool, which in combination with increased detector efficiency and reduced laboratory backgrounds, is particularly effective for measurements in which beam-induced backgrounds limit the experimental sensitivity. 

The improved background suppression, increased detection efficiency and new applications of the BGO as segmented detector at LUNA are discussed in next sections, in the context of the two scientific cases that they were designed for. A brief description of Monte Carlo simulations to study the detection efficiency and other characteristics of the detector is given beforehand, as this tool is used throughout this work.

\subsection{Simulations}
\label{sec:1.2}

\begin{figure}[htbp]
\centering
\setlength\lineskip{15pt}
    \includegraphics[width=0.85\textwidth]{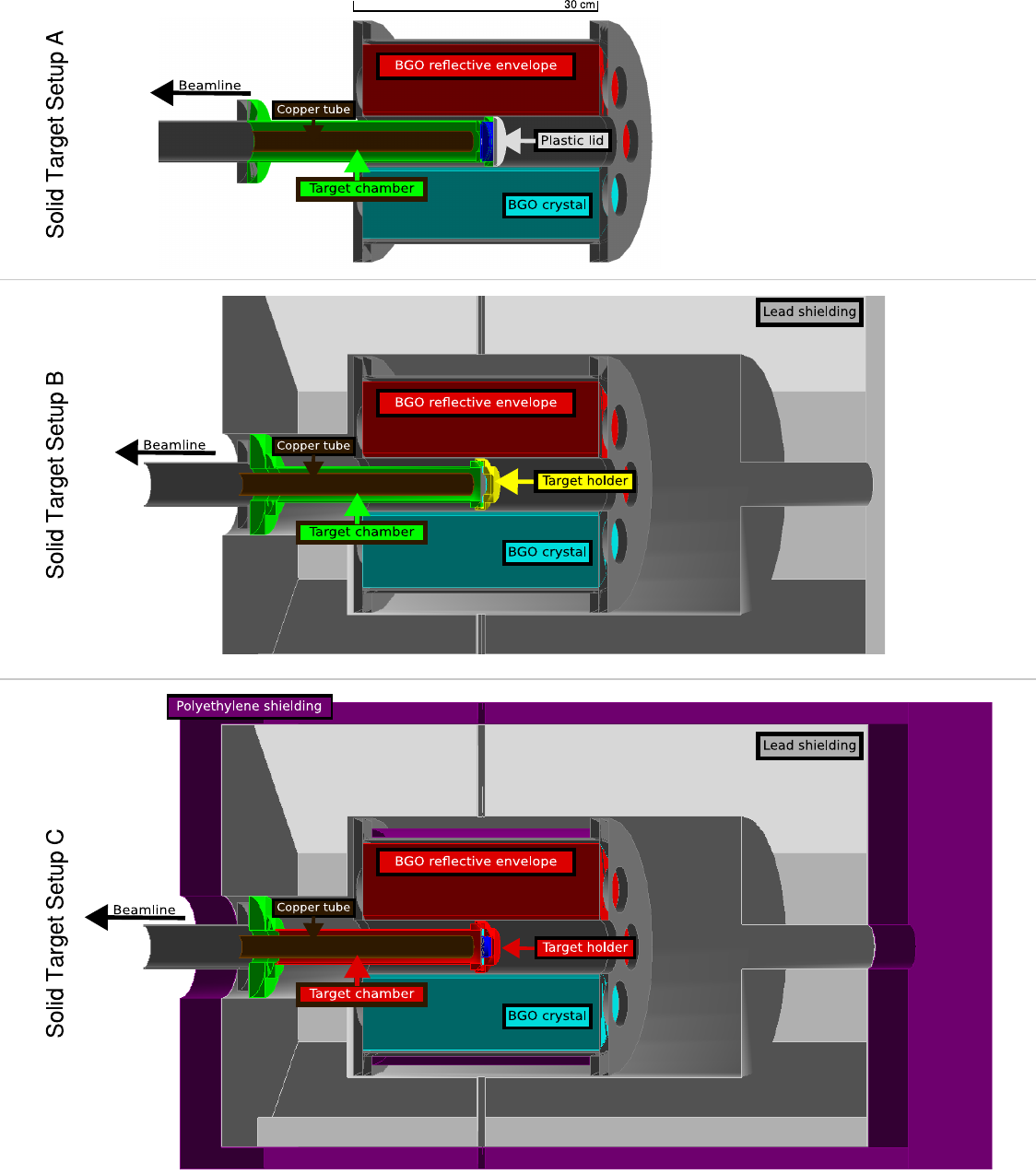}
   \caption{Cross section sketch of the three discussed setups. From top to bottom: Setup~A, B and C, see text for more details. Different colors correspond to different materials: green represents stainless steel, red aluminium, yellow brass, white plastic, grey lead, purple borated-Polyethylene, blue water, and cyan the BGO crystal.}
    \label{fig:Fig1-setups}
\end{figure}

A Monte Carlo particle transport simulation is a valuable tool to characterize the detector response, and explore the influence of different parameters of the setup.
For this purpose, a simulation based on the Geant4 toolkit \cite{Agostinelli_2003} was implemented on the basis of the adopted geometry for target, detector and shielding, as shown in Fig.~\ref{fig:Fig1-setups}. Details on the three setups in this figure are reported in next sections.


Once validated against calibration measurements, the simulation allows for a variety of applications in the data analysis. For example, the sum peak efficiency in the add-back spectrum can be obtained from the simulation, based on the known \gammaray{} cascades. Systematic effects, such as the influence of the beam spot position, or slight geometric asymmetries of the setup can be explored and taken into account for the analysis \citethesisboeltzig{}. Finally, for a given set of \gammaray{} cascades, the effect of applying gates in the add-back spectrum can be studied by virtue of the simulated data.

\begin{figure}[htbp]
    \centering
    \begin{minipage}{0.495\linewidth}
    \includegraphics[width=1\textwidth]{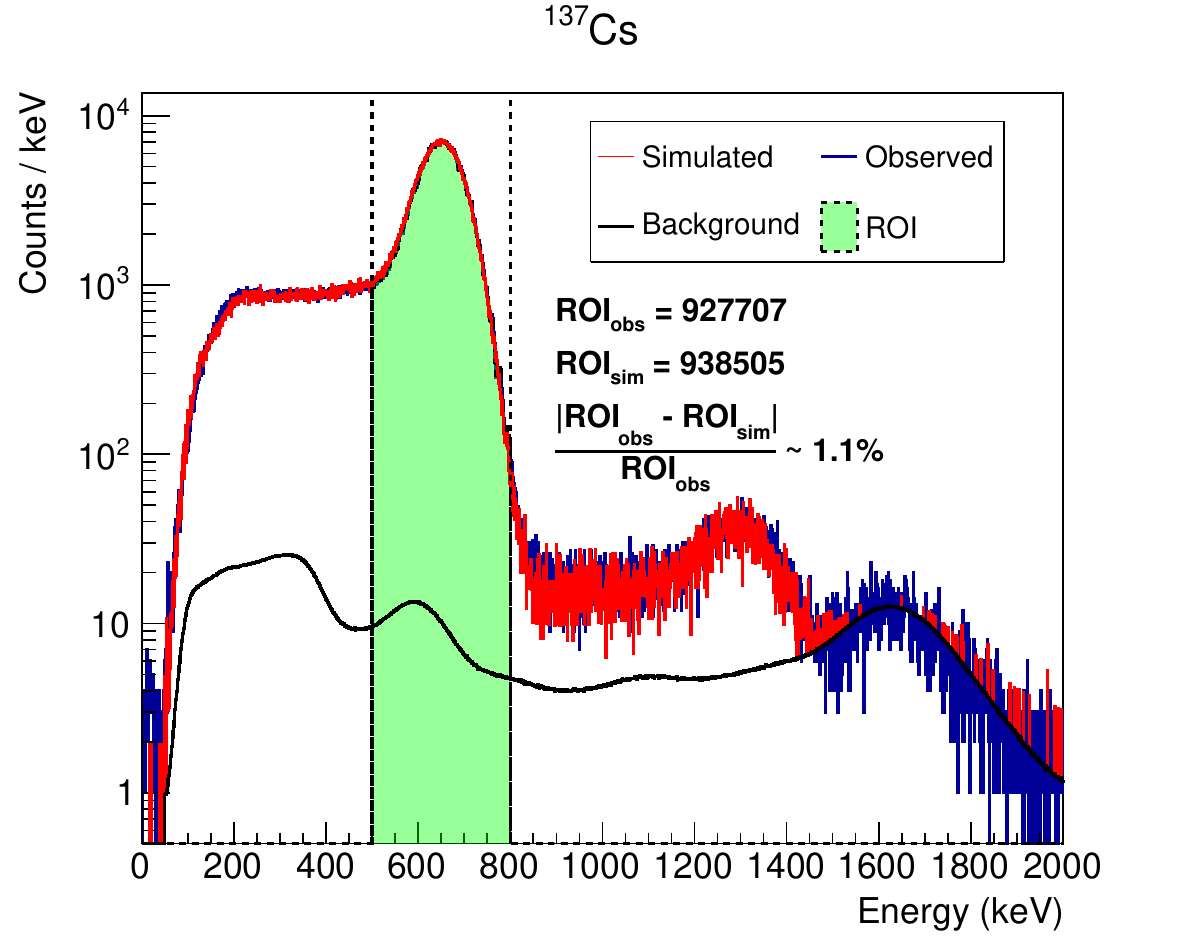}
    \end{minipage}
    \begin{minipage}{0.495\linewidth}
    \includegraphics[width=1\textwidth]{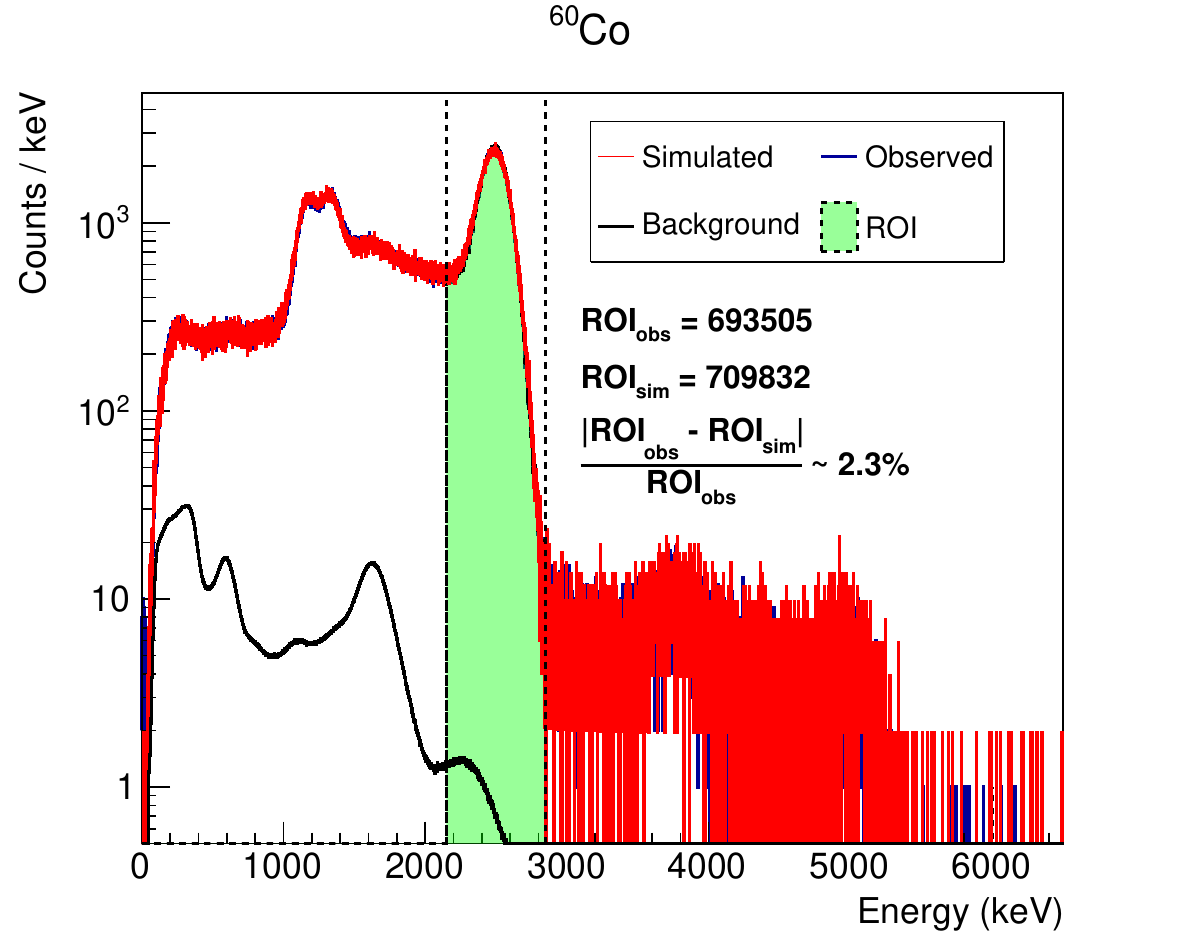}
    \end{minipage}
    \caption{Add-back \gammaray{} spectra of the $^{137}$Cs and $^{60}$Co calibration sources, comparing simulation with measurement. The agreement between the integrals in the respective regions of interest (ROI) is within \SI{3}{\percent}.}
    \label{fig:Fig3-sources}
\end{figure}

\begin{figure}[htbp]
    \centering
    \includegraphics[width=1\textwidth]{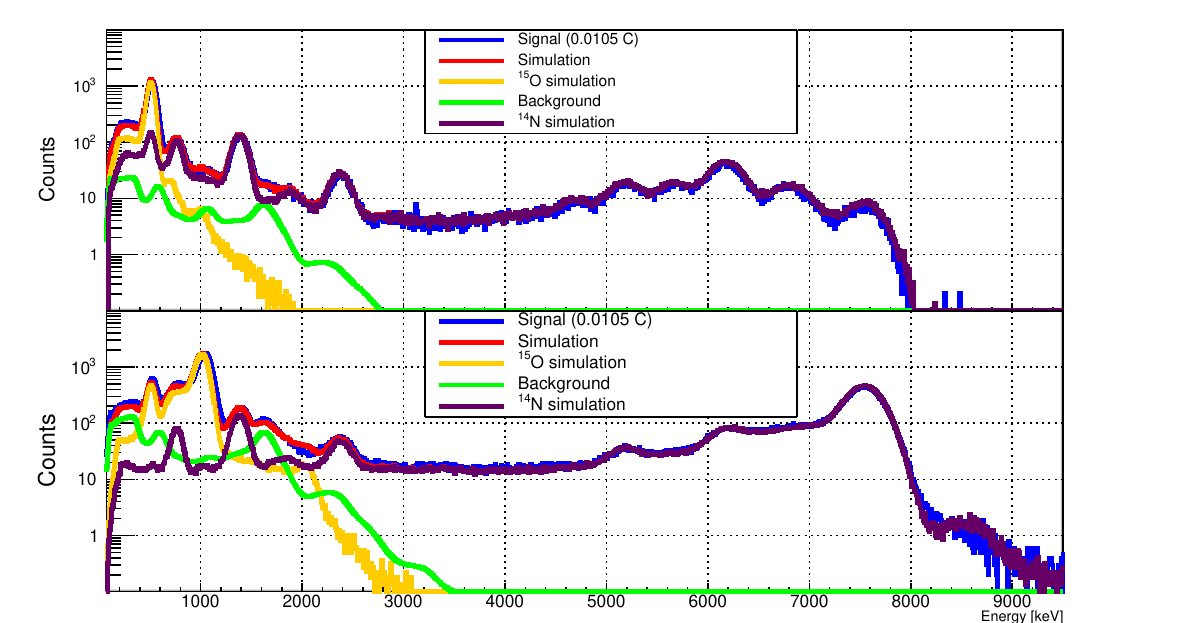}
    \caption{\gammaray{} spectra of the \reaction{14}{N}{\pg}{15}{O} $\Ecm = \SI{258}{keV}$ resonance, comparing simulation with measurement. Upper panel: single crystal spectrum (BGO~1). Lower panel: add-back spectrum. The agreement between the expected yield on the basis of~\cite{imbriani2005} and the simulations is within \SI{3}{\percent}, see text for details.}
    \label{fig:Fig2-simcompare}
\end{figure}

To validate simulations for each of the setups in Fig.~\ref{fig:Fig1-setups}, we compared the measured spectra of point-like \nuclide{60}{Co}, \nuclide{88}{Y} and 
\nuclide{137}{Cs} sources, which cover the low-energy part of the spectra. Good agreement was found between measurement and simulation, as illustrated in Fig.~\ref{fig:Fig3-sources}. The very well-known \Erescm{ = \SI{258}{keV}} resonance of
\reaction{14}{N}{\pg}{15}{O} ($\Ex = \SI{7556}{keV}$ \cite{selove1991}) \cite{imbriani2005}, allowed to extend the efficiency to higher energies. For simplicity we show the results of the validation procedure only for Setup~C.
Fine tuning the simulations focused at first on the analysis of the single BGO crystals, see top panel of Fig.~\ref{fig:Fig2-simcompare}. To reproduce the spectra on the whole energy range covered by the $\Ex = \SI{7556}{keV}$ de-excitation transitions, the energy resolution of each BGO crystal, random coincidences between two signals, and the decay of the \nuclide{15}{O} nucleus (\betaplus{} unstable with a half-life $\Thalf = \SI{2.037(2)}{\minute}$ \cite{selove1991}) were taken into account in the simulation, as well as the measured contribution of the laboratory background.
For the random summing (or pile-up) effect, a weighted sum of all possible combinations of signal sources was calculated, with weights determined by fitting the experimental pile-up peak. For example, Fig.~\ref{fig:Fig2-simcompare} shows pile-up between the prompt signal from \reaction{14}{N}{\pg}{15}{O} and the decay of \nuclide{15}{O} at around \SI{8.5}{MeV}, which is well-reproduced in the simulated spectrum.
The final agreement between simulated and measured spectra was within \SI{3}{\percent} for all crystals. In a next step, the comparison was extended to experimental and simulated add-back spectra, shown in the bottom panel of Fig.~\ref{fig:Fig2-simcompare}. Again, an agreement within about \SI{3}{\percent} was achieved. 
As the Monte Carlo simulations are used to evaluate the \gammaray{} detection efficiency of the detector, the discrepancy between the measured and the simulated efficiency observed in the calibration runs was taken as the systematic uncertainty on the efficiency determination. The statistical uncertainty in the calibration runs with the radioactive sources and the \reaction{14}{N}{\pg}{15}{O} reaction was well below \SI{1}{\percent}.


\section{Target and Shielding Setups with the BGO Detector at LUNA\label{sec:2}}

Since the commissioning of the BGO detector at LUNA, different experimental setups were developed for the individual reaction studies, guided by the main requirements of the targeted reaction. An overview is provided in Tab.~\ref{tab:Tab1-summary_setups}. The experiences with the previous iterations of the setup are reflected in its most recent upgrade. Here we describe the main modifications of the target chamber and the shielding setup for past and current experimental campaigns at LUNA and the following improvements on both efficiency and background reduction.

\begin{table}[htbp]
    \centering
    \begin{tabular}{c|c|c|c}
     Reaction  & $Q$-value [keV]&  Target Setup  &  Shielding \\
     \hline
     \reaction{2}{H}{\pg}{3}{He} \cite{casella_old_2002} & 5493  & Gas  & None\\
     \reaction{14}{N}{\pg}{15}{O}\cite{lemut_first_2006} & 7297 & Gas  & None \\
     \reaction{25}{Mg}{\pg}{26}{Al} \cite{strieder_25mgp_2012} & 6306 & A - Solid & None\\
     \reaction{18}{O}{\pg}{19}{F} \cite{best_18opg_2019} & 7994 & A - Solid & Pb \\
     \reaction{23}{Na}{\pg}{24}{Mg} \cite{boeltzig_23nap_2019} & 11693 & A - Solid & Pb\\
     \reaction{22}{Ne}{\pg}{23}{Na} \cite{Ferraro-2018PhRvL} & 8794 & Gas  & None\\
     \reaction{22}{Ne}{\ag}{26}{Mg}\cite{piatti_first_2022} & 10615 & Gas  & BPE (\SI{10}{cm}) thick\\
     \reaction{12}{C}{\pg}{13}{N} \cite{skowronski_epjwc_2022} & 1943 & B and C - Solid & Pb\\
     \reaction{13}{C}{\pg}{14}{N} \cite{skowronski_epjwc_2022} & 7551 & B - Solid & Pb\\
     \reaction{17}{O}{\pg}{18}{F} \cite{CianiPiatti_proceeding_2022} & 5607 & C - Solid & BPE + Pb + BPE (\SI{5}{cm})\\ 
    \end{tabular}
    \caption{Overview of the measurements performed with the BGO detector at LUNA. In the present paper we focus on target setups Solid A, B and C, see text and  Fig.\ref{fig:Fig1-setups} for details.}
    \label{tab:Tab1-summary_setups}
\end{table}

\subsection{Target Chambers and Efficiency\label{sec:2.2}}

We focus our discussion on three solid target setups here. Setup A was designed to measure the low-energy resonance of the \reaction{25}{Mg}{\pg}{26}{Al} reaction at \Erescm{ = \SI{92}{keV} (corresponding to $\Ex = \SI{6398}{keV}$) \cite{strieder_25mgp_2012}. A cross-sectional view of this setup is shown in the top panel of Fig.~\ref{fig:Fig1-setups}. The expected low count rate for this weak resonance emphasized the need for a large detection efficiency, \ie{}, the minimization of \gammaray{} absorption between the target and the detector. The cylindrical target chamber made of steel was designed to directly hold the targets, produced by evaporation on thin tantalum disks. This allowed for very little passive material between the target and the detector, at the expense of increased time and effort needed to mount or exchange the target. The same target setup has been successfully used for the study of \reaction{23}{Na}{\pg}{24}{Mg} \cite{boeltzig_23nap_2019}, and \reaction{18}{O}{\pg}{19}{F} \cite{best_18opg_2019}.

More recently, for the \reaction{13}{C}{\an}{16}{O} reaction measurement \cite{ciani_prl_2021} with a different detector, the need arose of very frequent target changes, and Setup~B was adopted: a brass target holder was designed to hold the tantalum target backings, with this holder being directly screwed onto the target chamber. This Setup~B allowed to minimize the time for target exchange when using at least two target holders. 
The practical advantages led to the use of the same setup also for the radiative proton capture measurements on carbon performed with the BGO detector.
The increased amount and density of passive materials, however, decreased the \gammaray{} detection efficiency by about \SI{14}{\percent} at $E_\gamma = \SI{1.332}{MeV}$, as shown by the comparison of simulation outputs for Setup~A and Setup~B in Fig.~\ref{fig:Fig3-oldneweffcomp}.
For the reactions \reaction{12,13}{C}{\pg}{13,14}{N}, relatively high reaction rates for all but the lowest accessed energies allowed for successful measurements in spite of the efficiency reduction.

To combine the advantages of both designs, ease of target exchange and high detection efficiency, the next revision of the experimental setup retained the chamber design of Setup~B, but introduced the use of aluminum both for the reaction chamber and the target holder, and further reducing the amount of material in the latter. The resulting Setup~C is shown in the bottom panel of Fig.~\ref{fig:Fig1-setups}. The main motivation for this setup was the first direct detection of the \Erescm{ = \SI{65}{keV}} resonance of the \reaction{17}{O}{\pg}{18}{F} reaction. For this we expected about \SI{0.09}{reactions/\hour} on the basis of literature data \cite{Buckner-2015}, assuming a beam current of \SI{100}{\mu A} and a \TantalumOxide{} target fully enriched in \nuclide{17}{O}. The revised setup was also decisive for the experimental campaign on \reaction{12}{C}{\pg}{13}{N} at low energies, allowing to extend the measurement down to $\Ep = \SI{80}{keV}$ corresponding to a cross section of the order of $\SI{1e-11}{b}$.
Compared to Setup~B, the efficiency of Setup~C is larger by \SI{24}{\percent} at $E_\gamma = \SI{1.332}{MeV}$ and is at least $\simeq \SI{18}{\percent}$ larger over the whole region up to \SI{6}{MeV}, see Fig.~\ref{fig:Fig3-oldneweffcomp}. Comparing with Setup~A, we observed an increased efficiency by $\simeq \SI{8}{\percent}$ at \SI{1.332}{MeV}.


\begin{figure}[htbp]
    \centering
    \includegraphics[width=1\textwidth]{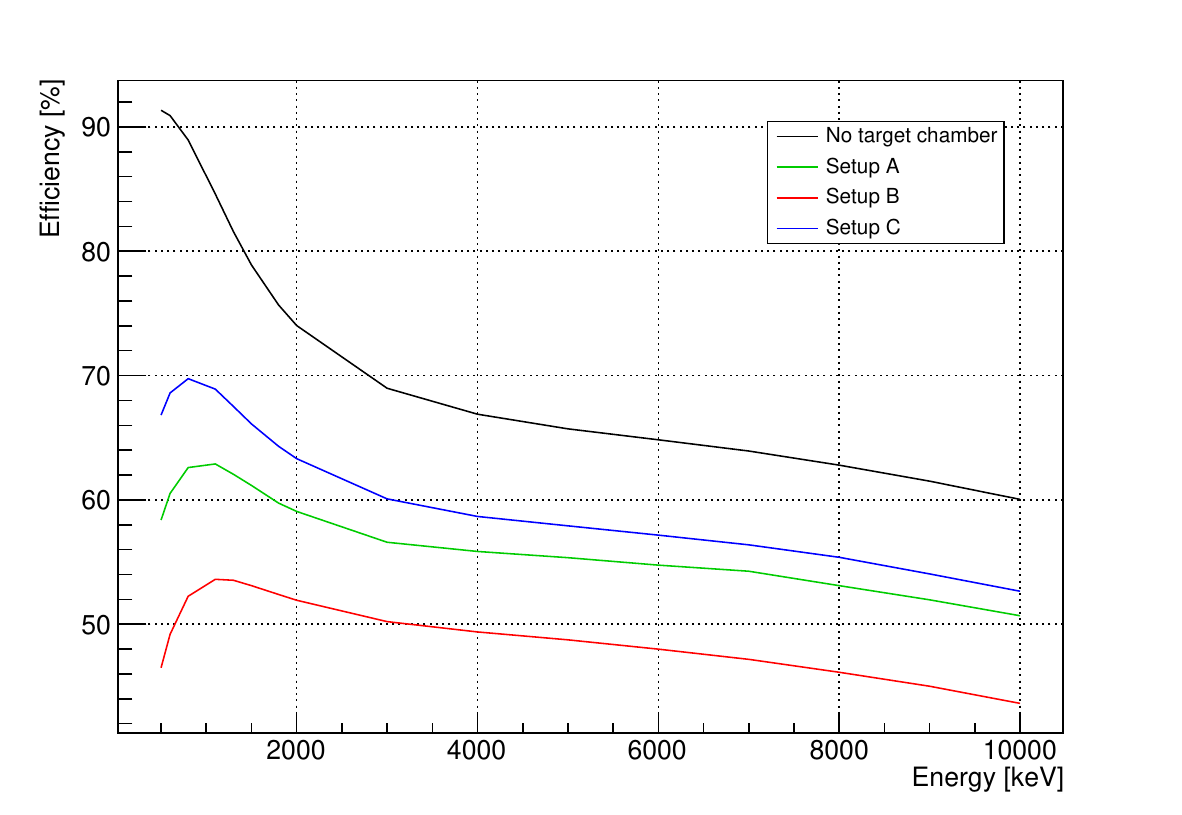}
    \caption{Add-back spectrum efficiency (sum peak efficiency) for single \gammaray{}s with energies from \SI{500}{keV} up to \SI{10}{MeV} obtained via simulations of Setup~A (green) Setup~B~(red) and Setup~C (blue). Simulations were implemented considering the complete target-detector-shielding geometry. Shown in black is a simulation of an idealized setup with no target chamber and target holder, thus with no \gammaray{} absorption.}
    \label{fig:Fig3-oldneweffcomp}
\end{figure}

\subsection{Shielding Evolution and Background Reduction\label{sec:2.5}}

Similarly to the target chamber setup, the BGO detector shielding evolved depending on the experimental necessities of the LUNA collaboration. Early experiments used no additional shielding to the BGO detector, both on solid targets \cite{strieder_25mgp_2012} and on gas targets \cite{casella_old_2002, lemut_first_2006, ferraro_high-efficiency_2018}.
Later experiments surrounded the BGO detector with \SI{10}{cm} of lead, to reduce environmental \gammaray{} background at low energies, its pile-up at medium energy, and some reduction of secondary backgrounds at higher energies. This shielding, first employed for the reaction \reaction{23}{Na}{\pg}{24}{Mg} ($Q = \SI{11693}{keV}$) \cite{boeltzig_23nap_2019}, was composed of two parts, mounted on rails to provide easy access to the reaction chamber to exchange the target when required. The effects of this shielding, together with a detailed model of the different background sources, are described in \cite{boeltzig_improved_2018}. 
This shielding resulted in a background reduction by two orders of magnitude at low energies, $E_\gamma < \SI{3}{MeV}$. In this energy region, the background is mainly due to environmental radioactivity (dominated by \nuclide{40}{K} and \nuclide{208}{Tl} peaks) and intrinsic radioactivity, which shows two main lines corresponding to the \SI{2340}{keV} and \SI{1633}{keV} states in \nuclide{207}{Pb} populated by the electron capture on \nuclide{207}{Bi}.

In the region of \SIrange{6}{18}{MeV} $\gamma$-energy, the remaining background is due to neutron-induced reactions on BGO materials, mainly Ge \cite{Bemmerer_et_al._2005, best-2016,boeltzig_improved_2018}.
For the measurement of the \reaction{22}{Ne}{\ag}{26}{Mg} ($Q = \SI{10615}{keV}$) on the gas target setup, a \SI{10}{cm} thick layer of borated (\SI{5}{\percent}) polyethylene (BPE) was added around the BGO detector, to absorb thermal neutrons and reduce the background above \SI{6}{MeV}.
This shielding reduced the counting rate in the region of interest (\SIrange{10}{11}{MeV}) by about a factor \num{3.4(3)} \cite{piatti_first_2022}.

This experience led to the optimization of the shielding in the region of interest for the \SI{65}{keV} resonance measurement of \reaction{17}{O}{\pg}{18}{F}. 
A three-layer shielding composed of (from outer to inner layer) \SI{5}{cm} of BPE, \SI{10}{cm} of lead and \SI{1}{cm} innermost BPE layer was built and installed (as illustrated in Fig.~\ref{fig:Fig1-setups}, setup C). 
This resulted in an overall background reduction by a factor \num{4.27(09)} in the region of interest, \SIrange{5200}{6200}{keV}, with respect to the lead shielding alone. A comparison of the measured backgrounds with the different shielding configuration is shown in Fig.~\ref{fig:Fig4-BackgroundComparison}.


\begin{figure}[htbp]
    \centering
    \includegraphics[width=1\textwidth]{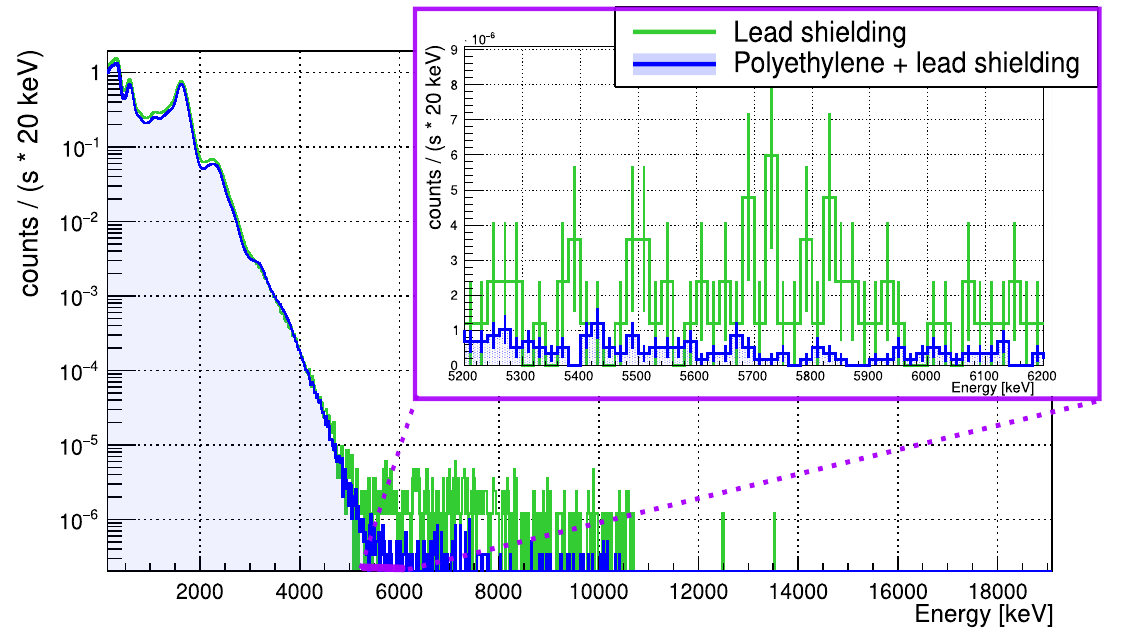}
    \caption{Measured background with Setup~B (black) and~C (blue). The inset shows the region of interest for the \Erescm = \SI{65}{keV} resonance of the  \reaction{17}{O}{\pg}{18}{F} reaction.}
    \label{fig:Fig4-BackgroundComparison}

\end{figure}


\section{\texorpdfstring
{Sensitivity study for \reaction{17}{O}{\pg}{18}{F}}
{Sensitivity study for 17O(p,gamma)18F}
}
\label{sec:3}

At the typical temperatures of shell hydrogen burning in Asymptotic Giant Branch (AGB) stars, \SIrange{0.03}{0.1}{GK}, corresponding to Gamow energies of \SIrange{35}{135}{keV}, the \shortreaction{17}{O}{\plusp} reaction rates are dominated by the \Erescm{ = \SI{65}{keV}} resonance ($\Ex = \SI{5672.57(32)}{keV}$ in \nuclide{18}{F} \cite{Tilley_1995}).
A recent direct measurement at LUNA has reported a resonance strength for the $\pa$ channel ($Q=\SI{1191}{keV}$) almost a factor of 2 larger than previously estimated  \cite{Bruno-2016PhRvL}, with significant impact on our knowledge of nucleosynthesis in intermediate-mass AGB stars \cite{Lugaro-2017NatAs, Straniero-2017A&A}.
For the \pg{} channel ($Q = \SI{5607}{keV}$), instead, only indirect measurements and following re-evaluations are reported in literature for the strength of this resonance. While $J^{\pi} = 1^-$ \cite{selove1991} and the partial widths $\Gamma_\gamma = \SI{0.44(2)}{eV}$ \cite{Buckner-2015} and $\Gamma_\alpha = \SI{130(5)}{eV}$ \cite{Mak-1980} are well constrained, the $\Gamma_\mathrm{p}$, calculated from the $\omega\gamma_{\pa}$, is the most uncertain quantity \cite{Bruno-2016PhRvL, Blackmon-1995, Sergi-2010, Hannam-1999}, with dramatic impact on the $\omega\gamma_{\pg}$ calculation.
The lowest recent estimate for the $\omega\gamma_{\pg} = \SI{1.6(3)e-11}{eV}$ \cite{Buckner-2015} is adopted in the following considerations.

LUNA combines an ideal site, long-standing experience and suitable tools for a high-sensitivity study of this resonance.
A feasibility study, partially reported in \cite{boeltzig_improved_2018} led to the final setup design (Setup~C, see Fig.~\ref{fig:Fig1-setups}), to perform at LUNA the first direct measurement of the $\SI{65}{keV}$ resonance strength. The target setup and its improved efficiency and background reduction were described in the previous section. The estimated sensitivity for Setup~C is illustrated in Fig.~\ref{fig:Fig5-17O_sensitivity}. The background level for this estimate was measured over \SI{68}{days}, leading to an average count rate of $\SI{2.6(3)e-8}{counts/(s \cdot 20\,keV})$.
Assuming a conservative \SI{200}{\mu A} current and a \TantalumOxide{} target with typical \nuclide{17}{O} enrichment of \SI{90}{\percent} \cite{Caciolli2012}, we simulated the add-back spectrum for a beam energy on top of the $\SI{65}{keV}$ resonance. The decay scheme of the state at $\Ex = \SI{5672}{keV}$ used in the simulation is listed in Tab.~\ref{tab:Tab2-17O_pg_cascades}, coincidences in the background measurement were reconstructed with the coincidence time window set to \SI{3.5}{\mu s} \cite{boeltzig_improved_2018}.


\begin{figure}[htbp]
    \centering
    \includegraphics[width=0.9\textwidth]{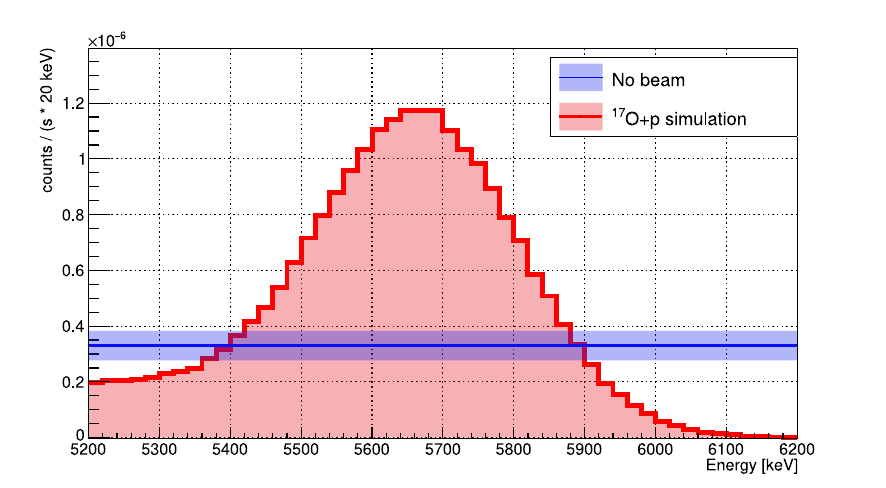}
    \caption{Measured environmental background rate when assuming a flat background in this region (blue), and simulated shape of the signal for the \reaction{17}{O}{\pg}{18}{F} reaction on the \SI{65}{keV} resonance (red). Assumptions of the simulation are described in the text. The blue band is the statistical error on the laboratory background.}
    \label{fig:Fig5-17O_sensitivity}
\end{figure}

The high sensitivity achieved with the present setup may be spoiled by target contaminants reacting with the beam. Contaminants could either be in the oxide film, or deeper in the backing. 
In order to monitor the beam-induced background, \TantalumOxide{} targets made with only Ultra Pure Water (UPW), thus containing a negligible amount of \nuclide{17}{O}, were irradiated.
To disentangle the \gammaray{}s from the beam-induced background and from the reaction of interest, we consider the \gammaray{} energies deposited in single crystals while gating on the sum energy peak of \reaction{17}{O}{\pg}{18}{F} in the add-back spectrum, as described in Sec.~\ref{sec:1.1}.
The segmentation of the detector, allows to distinctly identify the signature of the \reaction{17}{O}{\pg}{18}{F} reaction.
Gating on the ROI of the sum peak and looking at single crystal events, with multiplicity $> 1$, only events consistent with the cascade reported for the \SI{65}{keV} resonance are selected, Tab.~\ref{tab:Tab2-17O_pg_cascades}. This allows to effectively reject \gammaray{}s produced by beam-induced background emitting either single \gammaray{}s or cascades with different primaries and secondaries \gammaray{}s than those of the reaction of interest. In order to properly subtract the spurious coincidences due to beam induced background, which mimic the cascade of interest,  the same analysis is applied to the 65 keV resonance addback spectra acquired with UPW targets.


\begin{table}[htbp]
    \centering
    \begin{tabular}{|c|c|c|c|}
    \hline
    $\Erescm{}$ (keV) & $E_\mathrm{x}$ (keV)     &  $E_\mathrm{f}$ (keV)                          & $I_{\gamma}$ (\%) \\
   \hline
   {65} & 5672.57(32) \cite{Tilley_1995} & 3133.87(15) & 28.5(20) \\
   & & 3061.84 & 4.0(4) \\
   & & 2100.61 & 0.4(2) \\
   & & 1700.81 & 0.8(3) \\
   & & 1080.54 & 52(3) \\
   & & 1041.55 & 8.1(7) \\
   & & 0.0  & 6.2(4) \\
    \hline
    {183} & 5786(2) \cite{DiLeva2014} & 3791.49 & 4.5(4)\\ 
    & & 3358.2 & 2.3(3)\\ 
    & & 3133.87(15) & 4.3(4) \\
    & & 2523.35 & 5.5(6) \\
    & & 2100.61 & 11.8(8) \\
    & & 1080.54 & 40.8(7) \\
    & & 1041.55 & 3.4(4) \\
    & & 937 & 24.5(8) \\
    & & 0.0  & 2.9(4) \\
     \hline
  \end{tabular}
  \caption{Branching ratios for the excited states corresponding to the resonances at $\Erescm = \SI{65}{keV}$ and \SI{183}{keV} in \reaction{17}{O}{\pg}{18}{F} \cite{Tilley_1995, DiLeva2014}.}
  \label{tab:Tab2-17O_pg_cascades}
\end{table}

The performance of this technique was tested using the resonance \Erescm{ = \SI{183}{keV}} ($\Ex = \SI{5786(2)}{keV}$) of the \reaction{17}{O}{\pg}{18}{F} reaction \cite{DiLeva2014}. 
An experimental run on top of this resonance was performed by irradiating a \TantalumOxide{} target with a nominal isotopic enrichment of \SI{90}{\percent} in \nuclide{17}{O}. The add-back spectrum obtained in the measurement with the \nuclide{17}{O}-enriched target is compared in the top panel of Fig.~\ref{fig:Fig6-17O_gate} to a run taken under the same experimental conditions except for using a target made from UPW .
A gate in the ROI of the \SI{183}{keV} resonance sum peak was performed, both on signal and beam-induced background add-back spectra, and the contributions from all crystals were summed (Fig.~\ref{fig:Fig6-17O_gate}, bottom panel).  
In the $^{17}$O$\,+\,$p sum of the single spectra, the well known transitions corresponding to the $\Ex = \qty{5786}{keV}$ de-excitation are clearly visible and well in agreement with simulation based on literature branchings \ref{tab:Tab2-17O_pg_cascades}. 
In contrast, applying the gate to the UPW targets run does not highlight any structure.
Once a particular transition is selected, for example the main transition ($\rightarrow\SI{1080}{keV}$, $E_\gamma = \SI{4705}{keV}$, $I_\gamma = \SI{40.8(7)}{\percent}$ \cite{DiLeva2014}), the residual beam-induced background counts, \ie{}, the residual spurious coincidences that mimic the cascade of interest, is of $\SI{4.23e-4}{c/\mu C}$, leading to an increase of the signal to noise ratio by a factor of $8.6\,\pm\,1.9$ with respect to the sum peak case Fig.~\ref{fig:Fig6-17O_gate}.

This example demonstrates how the gate analysis allows to potentially identify and to cut off beam-induced background events, which populate the sum peak in the add-back spectrum, but do not match the signature of cascades of the target reaction \reaction{17}{O}{\pg}{18}{F}. The remaining beam-induced background in the UPW-target after the cut is due to random coincidences, and can be subtracted from the spectrum obtained with the \nuclide{17}{O}-enriched target.

For the case of the \Erescm{ = \SI{65}{keV}} resonance, a well known problem of tantalum is its ability to store hydrogen and deuterium \cite{Asakawa-2020JVSTB}. With the modest energy resolution of the BGO detector, the signature of a single \gammaray{} produced by the \reaction{2}{H}{\pg}{3}{He} reaction ($Q = \SI{5493}{keV}$) cannot be resolved from the signal of \reaction{17}{O}{\pg}{18}{F} ($Q = \SI{5607}{keV}$). The gate analysis is therefore a powerful tool in this case, to discriminate the \reaction{2}{H}{\pg}{3}{He} \gammaray{}s because of the more complex cascade of the \SI{65}{keV} resonance (Tab.~\ref{tab:Tab2-17O_pg_cascades}).


\begin{figure}[htbp]
    \centering
    \includegraphics[width=1\textwidth]{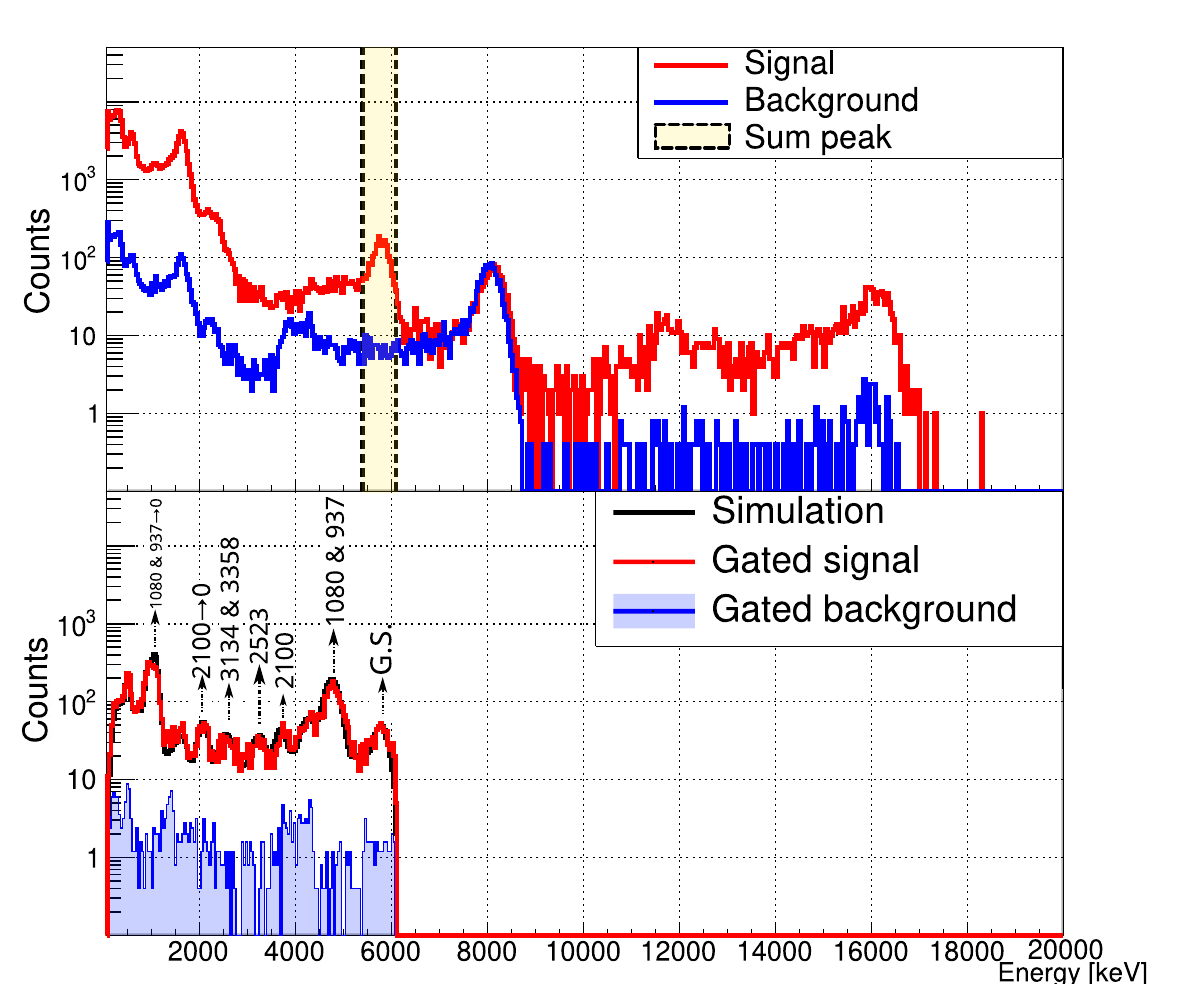}
    \caption{Upper panel: Add-back spectrum for proton irradiation at $\Ep = \SI{200}{keV}$ of a \nuclide{17}{O}-enriched target (red) and UPW-target (blue). The two spectra are normalized  to the \reaction{18}{O}{\pg}{19}{F} sum peak at about \SI{8}{MeV}. 
    Lower panel: Corresponding sum of single crystal spectra for both measurements after a cut of
    \SIrange{5400}{6094}{keV} in the sum spectrum, selecting the \reaction{17}{O}{\pg}{18}{F} signal. Distinct peaks corresponding to the transitions in \nuclide{18}{F} are visible in the red spectrum.}
    \label{fig:Fig6-17O_gate}
\end{figure}


\section{A New Application of a Segmented BGO Detector: Activation Counting}
\label{sec:4}


So far we focused on describing the classical application of the BGO summing detector underground: measurements of radiative capture reactions with high \Qvalue{}s, exploiting the ultra-low background in the corresponding region of interest. At lower energies, the BGO detector does not benefit as much from the deep-underground location, owing to the intrinsic radioactive background (substantially from \nuclide{207}{Bi}), as well as the remaining environmental radioactivity. Exploiting the unique signature following \betaplus{} decays, however, we present a new application of the detector to competitively measure such decays in spite of their low total \gammaray{} energy. 

Reaction products that are \betaplus{}-unstable decay by emission of e$^+$ particles, which soon after annihilate with an e$^-$. Counting the number of nuclei produced in the reaction by detecting the \SI{511}{keV} \gammaray{}s created in the e$^+$e$^-$ annihilation following their decay is a form of activation method~\cite{gyurky2019}.
The \SI{511}{keV} \gammaray{}s created in $e^+e^-$-annihilation are marked by a very distinct signature, as they are emitted in opposite directions from the point of annihilation. Because of the BGO segmentation and $4\pi$ geometry, this translates to a coincident detection of two \SI{511}{keV} signals in opposite crystals. Compared to off-site counting after irradiation as used in the previous LUNA measurements \cite{bemmerer_activation_2006, Scott2012}, both irradiation and annihilation counting can be performed \insitu{}.

This technique has been applied at LUNA to the \reaction{12}{C}{\pg}{13}{N} reaction study \cite{skowronski_epjwc_2022}. The emitted (single) prompt \gammaray{} has an energy of about  \SI{2}{MeV} ($Q = \SI{1943}{keV}$), thus sitting in a region affected by the intrinsic background of the detector. The presented activation technique is preferable since counting the number of \nuclide{13}{N} decays ($\Thalf = \SI{9.965(4)}{\minute}$ \cite{selove1991}) greatly improves the experimental sensitivity with this detector.
The study of the aforementioned reaction is still ongoing and will be discussed in a dedicated publication. In the following, we will focus on the reaction \reaction{14}{N}{\pg}{15}{O} (with \nuclide{15}{O} as unstable against \betaplus{} decay, $\Thalf = \SI{2.037(2)}{\minute}$ \cite{selove1991}), recently measured via activation technique at ATOMKI \cite{gyurky_activation_2019, gyurky_activation_2022}. We use this example to establish the technique, develop the according analysis procedures, and characterize the setup.

\subsection{Data Acquisition and Analysis}

Sputtered targets of TiN on Ta disks were used to study the \reaction{14}{N}{\pg}{15}{O} reaction.
Several irradiations (beam-on) were alternated with decay counting periods (beam-off). To help the search for \betaplus{} unstable contaminants, distinguished from the nuclide of interest only by their different half-lives, different lengths of the beam-off periods were used. In any case, no other such contaminants were observed for the presented case. Accurate knowledge of the beam current is crucial for this technique, and it is measured by a current integrator connected to the electrically insulated target chamber. The current as a function of time is recorded by using a digitizer, registering a pulse every \SI{1e-6}{C}. The number of counts was acquired in time-stamped list mode, allowing to obtain the charge rate in the offline analysis by counting the pulses from the integrator. The rates of the \SI{511}{keV} coincidences were derived from the list mode data by choosing an energy window of \SI{511(150)}{keV} and requiring that the coincidence events in opposite crystals were registered at most \SI{0.2}{\mu\second} apart in time~\cite{boeltzig_improved_2018}.

The data obtained, shown in Fig.~\ref{fig:Fig7-activ-fit}, describe the rate of events with an annihilation signature during irradiation and the subsequent exponential decay, as governed by the following relation:
\begin{equation}
  \frac{\mathrm{d}N}{\mathrm{d}t} = \eta \times Y \times R_\mathrm{p} - \lambda \times N(t), \label{eq:activ:de}
\end{equation}
where $Y$ is the reaction yield (reactions per incoming proton), $R_\mathrm{p}$ is the incoming proton rate, which can be derived from the recorded current, $\eta$ is the detection efficiency, which will be described in the next section, $\lambda$ is the decay constant of the decaying nucleus, and $N(t)$ is the number of nuclei present in the sample at a given time. The quantity that is directly observed from the data is the activity of the sample, $A(t) = \lambda \times N(t)$.

The observed data were iteratively fitted solving eq.~\ref{eq:activ:de}, and leaving $Y$ as the only free parameter. In addition, the Poisson likelihood was calculated at each iteration. In this way, the likelihood was maximized and the best-fit value was found for $Y$.

As eq.~\ref{eq:activ:de} is a differential equation, the initial condition $N(0)$ is required and it is calculated from the first data point, $A(0) = \lambda N(0)$. 
As residual activity from beam tuning or a preceding run may be present, $N(0)$ is not necessarily zero. The environmental background rate was calculated from a background measurement taken for \SI{20}{days}, and inserted in the fit as a nuisance parameter.
Another potential source of background are random coincidences caused by pair production or Compton scattering of prompt \gammaray{}s of beam-induced reactions. This contribution to the \SI{511}{keV} events is important only during the irradiation and was included in the final activity through a term $N_\mathrm{prompt} \times R_\mathrm{p}$, where $N_\mathrm{prompt}$ is the number of random coincidences per unit charge. This parameter is then left free to vary inside the fitting procedure. As an additional check, for the case of \reaction{14}{N}{\pg}{15}{O}, the best-fit $N_\mathrm{prompt}$ value was compared to the expected counts due to the prompt \gammaray{}s that can mimic the \SI{511}{keV} signals. For this purpose, the Geant4 simulation of the \nuclide{15}{O} \gammaray{} cascade at the resonance was used. This permitted to extract the probability of prompt \gammaray{}s creating random coincidence inside the BGO crystals. Finally the prompt \gammaray{} yields were analyzed, and compared to the results from the activation analysis. In Fig.~\ref{fig:Fig7-activ-fit} the fit for the \reaction{14}{N}{\pg}{15}{O} reaction at the \SI{259}{keV} resonances is shown. In Table~\ref{tab:Tab3-activ-results} the best-fit parameters are reported. An excellent agreement between prompt $\gamma$ and activation analysis was found. By taking the calculated background rate, $R_\mathrm{back}$, the estimated detection limit in terms of yield is \num{1e-18} with a beam current of \SI{400}{\mu A}.

\begin{table*}[htbp]
  \centering
  \begin{tabular}{|c|c|c|c|}
    \hline
    Method     & Yield  & $R_\mathrm{back}$ (s$^{-1}$) & $N_\mathrm{prompt}$ ($\mu$C$^{-1}$) \\
    \hline
    Activation & \num{6.64(5)e-12} & \num{0.0130(2)} & \num{0.13(2)}  \\
    Prompt     & \num{6.70(4)e-12} & & \\
    \hline
  \end{tabular}
  \caption{Overview of the results for \reaction{14}{N}{\pg}{15}{O} with the BGO detector, using activation or prompt $\gamma$ measurements. $R_\mathrm{back}$ is the rate of the random coincidences due to the environment.}
  \label{tab:Tab3-activ-results}
\end{table*}


\begin{figure}[htbp]
    \centering
    \includegraphics[width=0.7\textwidth]{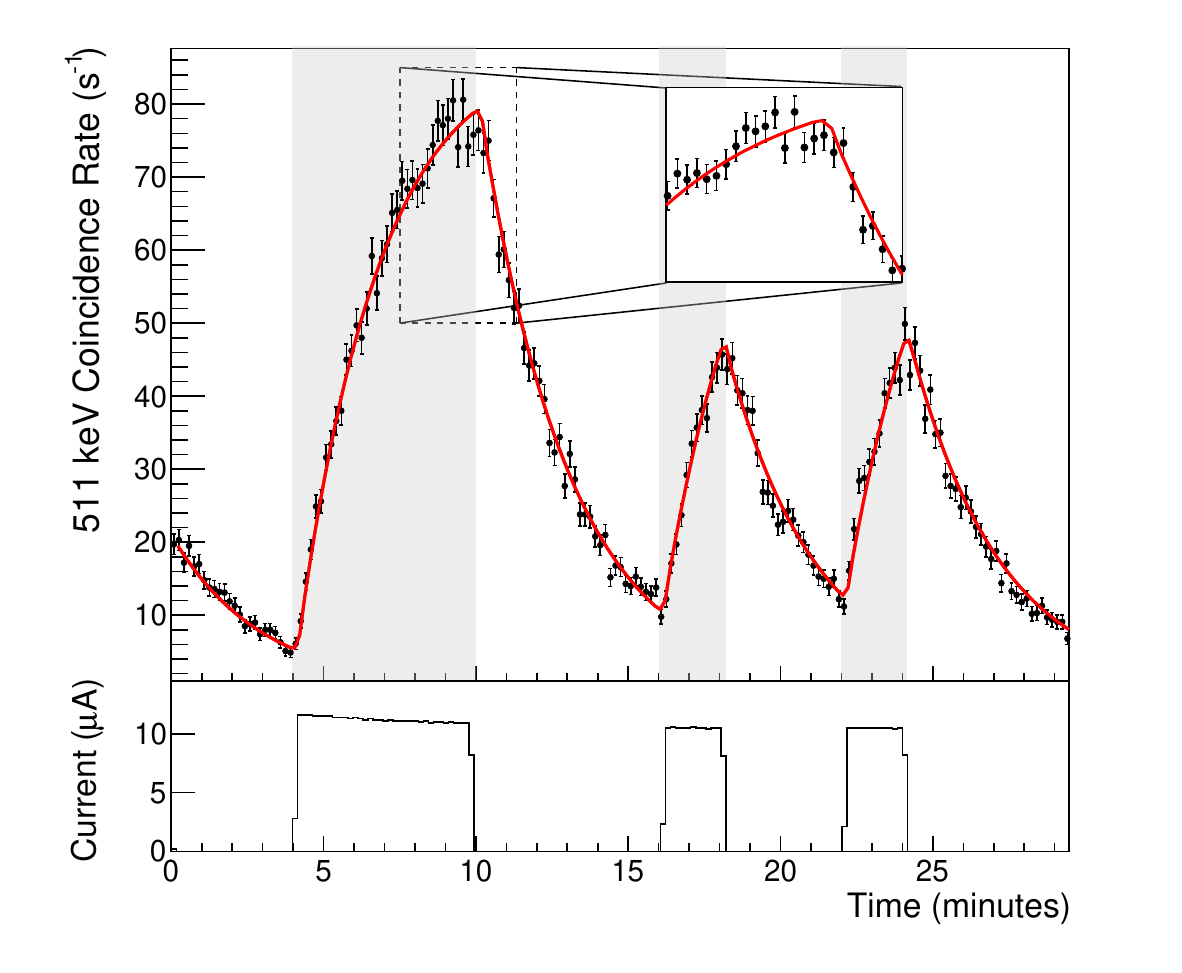}
    \caption{The upper panel shows the rate of the \SI{511}{keV} coincidences for the \nuclide{15}{O} activation on the \Erescm = \SI{258}{keV} resonance in \reaction{14}{N}{\pg}{15}{O}. The red line is the best-fit. The shaded areas represents the irradiation time slots. Different irradiation and counting periods were used to check for consistency and to search for any possible contaminant with a different half-life. The decaying activity in the first four minutes of the measurement without beam are the result of beam tuning before the start of data acquisition. The lower panel shows the current on target.}
    \label{fig:Fig7-activ-fit}
\end{figure}

\subsection{Efficiency}

The efficiency of detecting the annihilation signature for the \betaplus{} decay of a reaction product was obtained via Geant4 simulations of the detector setup, modelling the target geometry and tracking the \betaplus{} particles. The simulations were validated through comparison with radioactive source measurements (\nuclide{137}{Cs} and \nuclide{60}{Co}), as described in section \ref{sec:1.2}. The \betaplus{} decay of the \nuclide{15}{O} was simulated for a slightly off-center beam spot (as observed after dismounting the target), and at a depth in the target calculated with SRIM-2013 \cite{SRIM2003}.
The detection of two \SI{511}{keV} \gammaray{}s in coincidence was found to be quite sensitive to both the position of the beam spot (within \SI{2.4}{\percent}), and the depth of the decay inside the target (within \SI{1.0}{\percent}). This is due to the fact that, in contrast to the prompt \gammaray{}s, the \betaplus{} may travel some distance before annihilating and emitting the \SI{511}{keV} radiation. The simulated data were convoluted with the energy resolution of the detector, and the same coincidence energy window was applied to the simulation as for the measurement since the precise numerical value of the detection efficiency depends on it. Adding these contributions to the \SI{3}{\percent} agreement between the simulated and observed spectra resulted in a detection efficiency of \qty{21.4(9)}{\percent} for the \SI{511}{keV} coincidences in opposite BGO crystals.

To verify the validity of the calibration and of the method itself, an independent \exsitu{} measurement with an HPGe detector was performed using a \nuclide{12}{C} target initially irradiated \insitu{}. After irradiation, the activated sample was first counted \insitu{} with the BGO, then dismounted and installed in front of an HPGe detector (at a distance of about $\SI{14}{cm}$, with a Ta plate in front of the target to ensure positron annihilation), and finally re-mounted back in the initial position inside of the BGO detector for another counting period. Because of the different geometry, the HPGe detector counted single \SI{511}{keV} \gammaray{}s.
The HPGe detector was calibrated in efficiency by the use of radioactive sources, \nuclide{137}{Cs}, \nuclide{133}{Ba} and \nuclide{60}{Co}. The systematic uncertainty of the efficiency at \SI{511}{keV} is \SI{3}{\percent}, obtained from the efficiency curve minimization against the calibration sources.
The decay data from both the HPGe and the BGO, Fig.~\ref{fig:Fig8-activ-eff}, normalized for the respective detection efficiencies were fitted with an exponential decay with a half-life fixed to the literature value.
The good agreement found between the measurements with both detectors confirms the validity of the efficiency obtained from the simulations.

%
\begin{figure}[htbp]
    \centering
    \includegraphics[width=.7\textwidth]{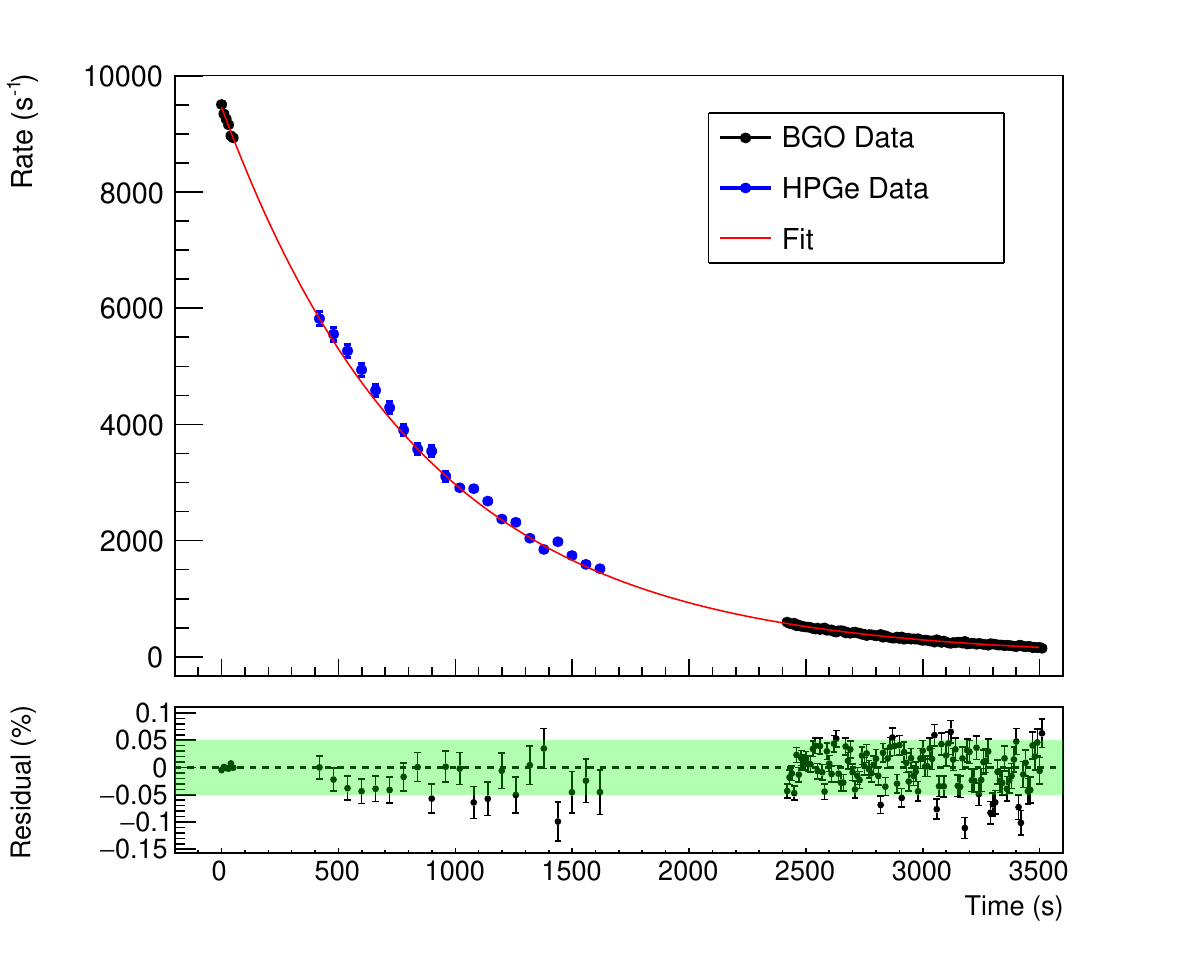}
    \caption{Results of the \exsitu{} calibration with an HPGe detector (blue), compared to \insitu{} measurements with the BGO detector (black) before and after. Only statistical uncertainty is plotted.}
    \label{fig:Fig8-activ-eff}
\end{figure}

\section{Conclusions}
\label{sec:5}

We reported the latest improvements of the detection setup with a BGO summing detector at LUNA, and their role for two distinct experimental studies: prompt \gammaray{} detection for \reaction{17}{O}{\pg}{18}{F}, and activation counting of \reaction{12}{C}{\pg}{13}{N}.

The achieved reduction of high-energy background rates, as well as the increase in detection efficiency, are crucial ingredients to directly detect the \Erescm{ = \SI{65}{keV}} resonance signal of the \reaction{17}{O}{\pg}{18}{F} reaction. Thanks to a three layered shielding, we achieved a background rate of \SI{1.8e-3}{c/(day \cdot keV)} in the region of interest, which corresponds to a reduction of about a factor of 4 with respect to the previous setup employing only a lead shielding.
Regarding the detection efficiency we reached a value of about \SI{60}{\percent} for a single \gammaray{} of \SI{6}{MeV}, which is the largest efficiency achieved with this detector so far at LUNA.
For a scenario with beam-induced backgrounds we have described an effective analysis based on time-coincidences to take advantage of the detector segmentation. Combined with the upgraded setup, the necessary sensitivity to detect \SI{0.09}{reactions/h} for the resonance in the relevant reaction \reaction{17}{O}{\pg}{18}{F} is within reach, and leads the way to future measurements at astrophysical energies.

Moreover, using the \reaction{14}{N}{\pg}{15}{O} reaction as a test case, we demonstrated a new mode of operation for the BGO detector by exploiting its segmentation for \insitu{} activation measurements of \betaplus{} unstable reaction products. Whilst the environmental background rate of the detector in this mode is not optimal (compared to the background level of an HPGe detector as reported in \cite{di_leva_underground_2014}), the advantage to perform irradiation and counting with the same setup and configuration allows for a high-efficiency detection of shorter-lived nuclides, most promising for half-lives of the order of seconds up to few hours.
One example is the lifetime of \nuclide{17}{F} ($\Thalf = \qty{64.49(16)}{s}$), as produced by the \reaction{16}{O}{\pg}{17}{F}, currently under study at LUNA, also via the new \insitu{} activation setup described here.


\ack

D. Ciccotti and the technical staff of the LNGS and INFN-Division of Padova and Naples mechanical workshops are gratefully acknowledged for their help.
Financial support by INFN,
the Italian Ministry of Education, University and Research (MIUR) through the ``Dipartimenti di eccellenza'' project ``Physics of the Universe'',
the European Union (ERC-CoG \emph{STARKEY}, no. 615604; ERC-StG \emph{SHADES}, no. 852016; and \emph{ChETEC-INFRA}, no. 101008324),
Deut\-sche For\-schungs\-ge\-mein\-schaft (DFG, BE~4100-4/1),
the Helm\-holtz Association (ERC-RA-0016),
the Hungarian National Research, Development and Innovation Office (NKFIH K134197),
the European Collaboration for Science and Technology (COST Action ChETEC, CA16117)
is gratefully acknowledged.
M.~A. , C.\,G.~B., T.~D., and R.\,S.~S.\ acknowledge funding by STFC UK (grant no. ST/L005824/1).{}

For the purpose of open access, the authors have applied a Creative Commons Attribution (CC BY) licence to any Author Accepted Manuscript version arising from this submission.

\section*{References}

\bibliographystyle{iopart-num.bst}
\bibliography{BGOtechnical2022} 

\providecommand{\newblock}{}
\begin{thebibliography}{10}
\expandafter\ifx\csname url\endcsname\relax
  \def\url#1{{\tt #1}}\fi
\expandafter\ifx\csname urlprefix\endcsname\relax\def\urlprefix{URL }\fi
\providecommand{\eprint}[2][]{\url{#2}}

\bibitem{aliotta_exploring_2022}
Aliotta M, Boeltzig A, Depalo R and Gy\"{u}rky G 2022 {\em Annual Review of
  Nuclear and Particle Science\/} {\bf 72} 177--204

\bibitem{Broggini-2018}
{Broggini} C, {Bemmerer} D, {Caciolli} A and {Trezzi} D 2018 {\em Progress in
  Particle and Nuclear Physics\/} {\bf 98} 55--84 (\textit{Preprint}
  \eprint{1707.07952})

\bibitem{Dombos_2022}
Dombos A~C, Robertson D, Simon A, Kadlecek T, Hanhardt M, G\"orres J, Couder M,
  Kelmar R, Olivas-Gomez O, Stech E, Strieder F and Wiescher M 2022 {\em Phys.
  Rev. Lett.\/} {\bf 128}(16) 162701

\bibitem{JUNA}
{Liu} W, {Li} Z, {He} J, {Tang} X, {Lian} G, {An} Z, {Chang} J, {Chen} H,
  {Chen} Q, {Chen} X, {Chen} Z, {Cui} B, {Du} X, {Fu} C, {Gan} L, {Guo} B, {He}
  G, {Heger} A, {Hou} S, {Huang} H, {Huang} N, {Jia} B, {Jiang} L, {Kubono} S,
  {Li} J, {Li} K, {Li} T, {Li} Y, {Lugaro} M, {Luo} X, {Ma} H, {Ma} S, {Mei} D,
  {Qian} Y, {Qin} J, {Ren} J, {Shen} Y, {Su} J, {Sun} L, {Tan} W, {Tanihata} I,
  {Wang} S, {Wang} P, {Wang} Y, {Wu} Q, {Xu} S, {Yan} S, {Yang} L, {Yang} Y,
  {Yu} X, {Yue} Q, {Zeng} S, {Zhang} H, {Zhang} H, {Zhang} L, {Zhang} N,
  {Zhang} Q, {Zhang} T, {Zhang} X, {Zhang} X, {Zhang} Z, {Zhao} W, {Zhao} Z and
  {Zhou} C 2016 {\em Science China Physics, Mechanics, and Astronomy\/} {\bf
  59} 5785

\bibitem{LUNA-MV}
Ferraro F, Ciani G~F, Boeltzig A, Cavanna F and Zavatarelli S 2021 {\em
  Frontiers in Astronomy and Space Sciences\/} {\bf 7} ISSN 2296-987X

\bibitem{Felsenkeller}
Bemmerer D, Cowan T~E, Domula A, D{\"o}ring T, Grieger M, Hammer S, Hensel T,
  H{\"u}binger L, Junghans A~R, Ludwig F, M{\"u}ller S~E, Reinicke S, Rimarzig
  B, Schmidt K, Schwengner R, St{\"o}ckel K, Sz{\"u}cs T, Turkat S, Wagner A,
  Wagner L and Zuber K 2019 {The new Felsenkeller 5 MV underground accelerator}
  {\em Solar Neutrinos: Proceedings of the 5$^{\rm th}$ International Solar
  Neutrino Conference\/} ed Meyer M and Zuber K (Singapore: {World Scientific})
  pp 249--263

\bibitem{Buckner-2015}
{Buckner} M~Q, {Iliadis} C, {Kelly} K~J, {Downen} L~N, {Champagne} A~E,
  {Cesaratto} J~M, {Howard} C and {Longland} R 2015 {\em Phys. Rev. C\/} {\bf
  91} 015812

\bibitem{Formicola-2003}
{Formicola} A, {Imbriani} G, {Junker} M, {Bemmerer} D, {Bonetti} R, {Broggini}
  C, {Casella} C, {Corvisiero} P, {Costantini} H, {Gervino} G, {Gustavino} C,
  {Lemut} A, {Prati} P, {Roca} V, {Rolfs} C, {Romano} M, {Sch{\"u}rmann} D,
  {Strieder} F, {Terrasi} F, {Trautvetter} H~P and {Zavatarelli} S 2003 {\em
  Nucl. Instr. Meth. Phys. Res. A\/} {\bf 507} 609--616

\bibitem{casella_new_2002}
Casella C, Costantini H, Lemut A, Limata B, Bemmerer D, Bonetti R, Broggini C,
  Campajola L, Cocconi P, Corvisiero P, Cruz J, D’Onofrio A, Formicola A,
  Fülöp Z, Gervino G, Gialanella L, Guglielmetti A, Gustavino C, Gyurky G,
  Loiano A, Imbriani G, Jesus A, Junker M, Musico P, Ordine A, Parodi F,
  Parolin M, Pinto J, Prati P, Ribeiro J, Roca V, Rogalla D, Rolfs C, Romano M,
  Rossi-Alvarez C, Rottura A, Schuemann F, Somorjai E, Strieder F, Terrasi F,
  Trautvetter H, Vomiero A and Zavatarelli S 2002 {\em Nucl. Instr. Meth. Phys.
  Res. A\/} {\bf 489} 160--169 ISSN 01689002

\bibitem{zhang_direct_2021}
Zhang L, Su J, He J, Wiescher M, deBoer R, Kahl D, Chen Y, Li X, Wang J, Zhang
  L, Cao F, Zhang H, Zhang Z, Jiao T, Sheng Y, Wang L, Song L, Jiang X, Li Z,
  Li E, Wang S, Lian G, Li Z, Tang X, Zhao H, Sun L, Wu Q, Li J, Cui B, Chen L,
  Ma R, Guo B, Xu S, Li J, Qi N, Sun W, Guo X, Zhang P, Chen Y, Zhou Y, Zhou J,
  He J, Shang C, Li M, Zhou X, Zhang Y, Zhang F, Hu Z, Xu H, Chen J and Liu W
  2021 {\em Phys. Rev. Lett.\/} {\bf 127} 152702 ISSN 0031-9007, 1079-7114

\bibitem{Zhang_19F_pg_2022}
Zhang L, He J, deBoer R~J, Wiescher M, Heger A, Kahl D, Su J, Odell D, Chen Y,
  Li X, Wang J, Zhang L, Cao F, Zhang H, Zhang Z, Jiang X, Wang L, Li Z, Song
  L, Zhao H, Sun L, Wu Q, Li J, Cui B, Chen L, Ma R, Li E, Lian G, Sheng Y, Li
  Z, Guo B, Zhou X, Zhang Y, Xu H, Cheng J and Liu W 2022 {\em Nature\/} {\bf
  610} 656--660 ISSN 1476-4687

\bibitem{HECTOR-CASPAR}
Olivas-Gomez O, Simon A, Robertson D, Dombos A~C, Strieder F, Kadlecek T,
  Hanhardt M, Kelmar R, Couder M, Görres J, Stech E and Wiescher M 2022 {\em
  EPJ A\/} {\bf 58} 57 ISSN 1434-6001, 1434-601X

\bibitem{Shahina_direct_2022}
Shahina, G\"orres J, Robertson D, Couder M, Gomez O, Gula A, Hanhardt M,
  Kadlecek T, Kelmar R, Scholz P, Simon A, Stech E, Strieder F and Wiescher M
  2022 {\em Phys. Rev. C\/} {\bf 106}(2) 025805

\bibitem{casella_old_2002}
{Casella} C, {Costantini} H, {Lemut} A, {Limata} B, {Bonetti} R, {Broggini} C,
  {Campajola} L, {Corvisiero} P, {Cruz} J, {D'Onofrio} A, {Formicola} A,
  {F{\"u}l{\"o}p} Z, {Gervino} G, {Gialanella} L, {Guglielmetti} A, {Gustavino}
  C, {Gyurky} G, {Imbriani} G, {Jesus} A~P, {Junker} M, {Ordine} A, {Pinto}
  J~V, {Prati} P, {Ribeiro} J~P, {Roca} V, {Rogalla} D, {Rolfs} C, {Romano} M,
  {Rossi-Alvarez} C, {Schuemann} F, {Somorjai} E, {Straniero} O, {Strieder} F,
  {Terrasi} F, {Trautvetter} H~P, {Zavatarelli} S and {LUNA Collaboration} 2002
  {\em Nucl. Phys. A\/} {\bf 706} 203--216

\bibitem{lemut_first_2006}
Lemut A, Bemmerer D, Confortola F, Bonetti R, Broggini C, Corvisiero P,
  Costantini H, Cruz J, Formicola A, Fülöp Z, Gervino G, Guglielmetti A,
  Gustavino C, Gyürky G, Imbriani G, Jesus A, Junker M, Limata B, Menegazzo R,
  Prati P, Roca V, Rogalla D, Rolfs C, Romano M, Rossi~Alvarez C, Schümann F,
  Somorjai E, Straniero O, Strieder F, Terrasi F and Trautvetter H 2006 {\em
  Phys. Lett. B\/} {\bf 634} 483--487 ISSN 03702693

\bibitem{bemmerer_low_2006}
Bemmerer D, Confortola F, Lemut A, Bonetti R, Broggini C, Corvisiero P,
  Costantini H, Cruz J, Formicola A, Fülöp Z, Gervino G, Guglielmetti A,
  Gustavino C, Gyürky G, Imbriani G, Jesus A, Junker M, Limata B, Menegazzo R,
  Prati P, Roca V, Rolfs C, Rogalla D, Romano M, Rossi-Alvarez C, Schümann F,
  Somorjai E, Straniero O, Strieder F, Terrasi F and Trautvetter H 2006 {\em
  Nucl. Phys. A\/} {\bf 779} 297--317 ISSN 03759474

\bibitem{strieder_25mgp_2012}
{Strieder} F, {Limata} B, {Formicola} A, {Imbriani} G, {Junker} M, {Bemmerer}
  D, {Best} A, {Broggini} C, {Caciolli} A, {Corvisiero} P, {Costantini} H,
  {DiLeva} A, {Elekes} Z, {F{\"u}l{\"o}p} Z, {Gervino} G, {Guglielmetti} A,
  {Gustavino} C, {Gy{\"u}rky} G, {Lemut} A, {Marta} M, {Mazzocchi} C,
  {Menegazzo} R, {Prati} P, {Roca} V, {Rolfs} C, {Rossi Alvarez} C, {Somorjai}
  E, {Straniero} O, {Terrasi} F and {Trautvetter} H~P 2012 {\em Phys. Lett.
  B\/} {\bf 707} 60--65

\bibitem{su_first_2021}
Su J, Zhang H, Li Z, Ventura P, Li Y, Li E, Chen C, Shen Y, Lian G, Guo B, Li
  X, Zhang L, He J, Sheng Y, Chen Y, Wang L, Zhang L, Cao F, Nan W, Nan W, Li
  G, Song N, Cui B, Chen L, Ma R, Zhang Z, Jiao T, Gao B, Tang X, Wu Q, Li J,
  Sun L, Wang S, Yan S, Liao J, Wang Y, Zeng S, Nan D, Fan Q, Qi N, Sun W, Guo
  X, Zhang P, Chen Y, Zhou Y, Zhou J, He J, Shang C, Li M and Liu W 2021 {\em
  Science Bulletin\/}  S2095927321006745 ISSN 20959273

\bibitem{piatti_first_2022}
Piatti D, Masha E, Aliotta M, Balibrea-Correa J, Barile F, Bemmerer D, Best A,
  Boeltzig A, Broggini C, Bruno C~G, Caciolli A, Cavanna F, Chillery T, Ciani
  G~F, Compagnucci A, Corvisiero P, Csedreki L, Davinson T, Depalo R, Leva A~d,
  Elekes Z, Ferraro F, Fiore E~M, Formicola A, F{\"u}l{\"o}p Z, Gervino G,
  Guglielmetti A, Gustavino C, Gy{\"u}rky G, Imbriani G, Junker M, Lugaro M,
  Marigo P, Menegazzo R, Mossa V, Pantaleo F~R, Paticchio V, Perrino R, Prati
  P, Rapagnani D, Schiavulli L, Skowronski J, St{\"o}ckel K, Straniero O,
  Sz{\"u}cs T, Tak{\'a}cs M~P and Zavatarelli S 2022 {\em EPJ A\/} {\bf 58} 194
  ISSN 1434-601X

\bibitem{boeltzig_improved_2018}
Boeltzig A, Best A, Imbriani G, Junker M, Aliotta M, Bemmerer D, Broggini C,
  Bruno C~G, Buompane R, Caciolli A, Cavanna F, Chillery T, Ciani G~F,
  Corvisiero P, Csedreki L, Davinson T, deBoer R~J, Depalo R, Leva A~D, Elekes
  Z, Ferraro F, Fiore E~M, Formicola A, Fülöp Z, Gervino G, Guglielmetti A,
  Gustavino C, Gyürky G, Kochanek I, Menegazzo R, Mossa V, Pantaleo F~R,
  Paticchio V, Perrino R, Piatti D, Prati P, Schiavulli L, Stöckel K,
  Straniero O, Strieder F, Szücs T, Takács M~P, Trezzi D, Wiescher M and
  Zavatarelli S 2018 {\em J. Phys. G\/} {\bf 45} 025203 ISSN 0954-3899,
  1361-6471

\bibitem{ferraro_high-efficiency_2018}
Ferraro F, Takács M~P, Piatti D, Mossa V, Aliotta M, Bemmerer D, Best A,
  Boeltzig A, Broggini C, Bruno C~G, Caciolli A, Cavanna F, Chillery T, Ciani
  G~F, Corvisiero P, Csedreki L, Davinson T, Depalo R, D’Erasmo G, Di~Leva A,
  Elekes Z, Fiore E~M, Formicola A, Fülöp Z, Gervino G, Guglielmetti A,
  Gustavino C, Gyürky G, Imbriani G, Junker M, Kochanek I, Lugaro M, Marcucci
  L~E, Marigo P, Menegazzo R, Pantaleo F~R, Paticchio V, Perrino R, Prati P,
  Schiavulli L, Stöckel K, Straniero O, Szücs T, Trezzi D and Zavatarelli S
  2018 {\em EPJ A\/} {\bf 54} 44 ISSN 1434-6001, 1434-601X

\bibitem{Cavanna-2015}
{Cavanna} F, {Depalo} R, {Aliotta} M, {Anders} M, {Bemmerer} D, {Best} A,
  {Boeltzig} A, {Broggini} C, {Bruno} C~G, {Caciolli} A, {Corvisiero} P,
  {Davinson} T, {di Leva} A, {Elekes} Z, {Ferraro} F, {Formicola} A,
  {F{\"u}l{\"o}p} Z, {Gervino} G, {Guglielmetti} A, {Gustavino} C, {Gy{\"u}rky}
  G, {Imbriani} G, {Junker} M, {Menegazzo} R, {Mossa} V, {Pantaleo} F~R,
  {Prati} P, {Scott} D~A, {Somorjai} E, {Straniero} O, {Strieder} F,
  {Sz{\"u}cs} T, {Tak{\'a}cs} M~P, {Trezzi} D and {LUNA Collaboration} 2015
  {\em Phys. Rev. Lett.\/} {\bf 115} 252501 (\textit{Preprint}
  \eprint{1511.05329})

\bibitem{Agostinelli_2003}
{Agostinelli} S, {Allison} J, {Amako} K, {Apostolakis} J, {Araujo} H, {Arce} P,
  {Asai} M, {Axen} D, {Banerjee} S, {Barrand} G, {Behner} F, {Bellagamba} L,
  {Boudreau} J, {Broglia} L, {Brunengo} A, {Burkhardt} H, {Chauvie} S, {Chuma}
  J, {Chytracek} R, {Cooperman} G, {Cosmo} G, {Degtyarenko} P, {Dell'Acqua} A,
  {Depaola} G, {Dietrich} D, {Enami} R, {Feliciello} A, {Ferguson} C,
  {Fesefeldt} H, {Folger} G, {Foppiano} F, {Forti} A, {Garelli} S, {Giani} S,
  {Giannitrapani} R, {Gibin} D, {G{\'o}mez Cadenas} J~J, {Gonz{\'a}lez} I,
  {Gracia Abril} G, {Greeniaus} G, {Greiner} W, {Grichine} V, {Grossheim} A,
  {Guatelli} S, {Gumplinger} P, {Hamatsu} R, {Hashimoto} K, {Hasui} H,
  {Heikkinen} A, {Howard} A, {Ivanchenko} V, {Johnson} A, {Jones} F~W,
  {Kallenbach} J, {Kanaya} N, {Kawabata} M, {Kawabata} Y, {Kawaguti} M,
  {Kelner} S, {Kent} P, {Kimura} A, {Kodama} T, {Kokoulin} R, {Kossov} M,
  {Kurashige} H, {Lamanna} E, {Lamp{\'e}n} T, {Lara} V, {Lefebure} V, {Lei} F,
  {Liendl} M, {Lockman} W, {Longo} F, {Magni} S, {Maire} M, {Medernach} E,
  {Minamimoto} K, {Mora de Freitas} P, {Morita} Y, {Murakami} K, {Nagamatu} M,
  {Nartallo} R, {Nieminen} P, {Nishimura} T, {Ohtsubo} K, {Okamura} M,
  {O'Neale} S, {Oohata} Y, {Paech} K, {Perl} J, {Pfeiffer} A, {Pia} M~G,
  {Ranjard} F, {Rybin} A, {Sadilov} S, {Di Salvo} E, {Santin} G, {Sasaki} T,
  {Savvas} N, {Sawada} Y, {Scherer} S, {Sei} S, {Sirotenko} V, {Smith} D,
  {Starkov} N, {Stoecker} H, {Sulkimo} J, {Takahata} M, {Tanaka} S,
  {Tcherniaev} E, {Safai Tehrani} E, {Tropeano} M, {Truscott} P, {Uno} H,
  {Urban} L, {Urban} P, {Verderi} M, {Walkden} A, {Wander} W, {Weber} H,
  {Wellisch} J~P, {Wenaus} T, {Williams} D~C, {Wright} D, {Yamada} T, {Yoshida}
  H, {Zschiesche} D and {G EANT4 Collaboration} 2003 {\em Nucl. Instr. Meth.
  Phys. Res. A\/} {\bf 506} 250--303

\bibitem{thesis:boeltzig_unblinded}
Boeltzig A 2016 {\em Direct measurements of the $^{23}Na(p,\gamma)^{24}Mg$
  cross section at stellar energies\/} {PhD thesis} Gran Sasso Science
  Institute

\bibitem{selove1991}
Ajzenberg-Selove F 1991 {\em Nucl. Phys. A\/} {\bf 523} 1--196 ISSN 0375-9474

\bibitem{imbriani2005}
{Imbriani} G, {Costantini} H, {Formicola} A, {Vomiero} A, {Angulo} C,
  {Bemmerer} D, {Bonetti} R, {Broggini} C, {Confortola} F, {Corvisiero} P,
  {Cruz} J, {Descouvemont} P, {F{\"u}l{\"o}p} Z, {Gervino} G, {Guglielmetti} A,
  {Gustavino} C, {Gy{\"u}rky} G, {Jesus} A~P, {Junker} M, {Klug} J~N, {Lemut}
  A, {Menegazzo} R, {Prati} P, {Roca} V, {Rolfs} C, {Romano} M, {Rossi-Alvarez}
  C, {Sch{\"u}mann} F, {Sch{\"u}rmann} D, {Somorjai} E, {Straniero} O,
  {Strieder} F, {Terrasi} F and {Trautvetter} H~P ({LUNA Collaboration}) 2005
  {\em EPJ A\/} {\bf 25} 455--466

\bibitem{best_18opg_2019}
{Best} A, {Pantaleo} F~R, {Boeltzig} A, {Imbriani} G, {Aliotta} M,
  {Balibrea-Correa} J, {Bemmerer} D, {Broggini} C, {Bruno} C~G, {Buompane} R,
  {Caciolli} A, {Cavanna} F, {Chillery} T, {Ciani} G~F, {Corvisiero} P,
  {Csedreki} L, {Davinson} T, {deBoer} R~J, {Depalo} R, {Di Leva} A, {Elekes}
  Z, {Ferraro} F, {Fiore} E~M, {Formicola} A, {F{\"u}l{\"o}p} Z, {Gervino} G,
  {Guglielmetti} A, {Gustavino} C, {Gy{\"u}rky} G, {Junker} M, {Kochanek} I,
  {Lugaro} M, {Marigo} P, {Menegazzo} R, {Mossa} V, {Paticchio} V, {Perrino} R,
  {Piatti} D, {Prati} P, {Schiavulli} L, {St{\"o}ckel} K, {Straniero} O,
  {Strieder} F, {Sz{\"u}cs} T, {Tak{\'a}cs} M~P, {Trezzi} D, {Wiescher} M and
  {Zavatarelli} S 2019 {\em Phys. Lett. B\/} {\bf 797} 134900

\bibitem{boeltzig_23nap_2019}
{Boeltzig} A, {Best} A, {Pantaleo} F~R, {Imbriani} G, {Junker} M, {Aliotta} M,
  {Balibrea-Correa} J, {Bemmerer} D, {Broggini} C, {Bruno} C~G, {Buompane} R,
  {Caciolli} A, {Cavanna} F, {Chillery} T, {Ciani} G~F, {Corvisiero} P,
  {Csedreki} L, {Davinson} T, {deBoer} R~J, {Depalo} R, {Di Leva} A, {Elekes}
  Z, {Ferraro} F, {Fiore} E~M, {Formicola} A, {F{\"u}l{\"o}p} Z, {Gervino} G,
  {Guglielmetti} A, {Gustavino} C, {Gy{\"u}rky} G, {Kochanek} I, {Lugaro} M,
  {Marigo} P, {Menegazzo} R, {Mossa} V, {Munnik} F, {Paticchio} V, {Perrino} R,
  {Piatti} D, {Prati} P, {Schiavulli} L, {St{\"o}ckel} K, {Straniero} O,
  {Strieder} F, {Sz{\"u}cs} T, {Tak{\'a}cs} M~P, {Trezzi} D, {Wiescher} M and
  {Zavatarelli} S 2019 {\em Phys. Lett. B\/} {\bf 795} 122--128

\bibitem{Ferraro-2018PhRvL}
{Ferraro} F, {Tak{\'a}cs} M~P, {Piatti} D, {Cavanna} F, {Depalo} R, {Aliotta}
  M, {Bemmerer} D, {Best} A, {Boeltzig} A, {Broggini} C, {Bruno} C~G,
  {Caciolli} A, {Chillery} T, {Ciani} G~F, {Corvisiero} P, {Davinson} T,
  {D'Erasmo} G, {Di Leva} A, {Elekes} Z, {Fiore} E~M, {Formicola} A,
  {F{\"u}l{\"o}p} Z, {Gervino} G, {Guglielmetti} A, {Gustavino} C, {Gy{\"u}rky}
  G, {Imbriani} G, {Junker} M, {Karakas} A, {Kochanek} I, {Lugaro} M, {Marigo}
  P, {Menegazzo} R, {Mossa} V, {Pantaleo} F~R, {Paticchio} V, {Perrino} R,
  {Prati} P, {Schiavulli} L, {St{\"o}ckel} K, {Straniero} O, {Sz{\"u}cs} T,
  {Trezzi} D, {Zavatarelli} S and {LUNA Collaboration} 2018 {\em Phys. Rev.
  Lett.\/} {\bf 121} 172701 (\textit{Preprint} \eprint{1810.01628})

\bibitem{skowronski_epjwc_2022}
Skowronski J ({LUNA}) 2022 {\em European Physical Journal Web of Conferences\/}
  {\bf 260} 11008 ISSN 2100-014X

\bibitem{CianiPiatti_proceeding_2022}
{Ciani} G~F, {Piatti} D and {Gesu{\`e}} R~M 2022 {\em European Physical Journal
  Web of Conferences\/} {\bf 260} 11003

\bibitem{ciani_prl_2021}
{Ciani} G~F, {Csedreki} L, {Rapagnani} D, {Aliotta} M, {Balibrea-Correa} J,
  {Barile} F, {Bemmerer} D, {Best} A, {Boeltzig} A, {Broggini} C, {Bruno} C~G,
  {Caciolli} A, {Cavanna} F, {Chillery} T, {Colombetti} P, {Corvisiero} P,
  {Cristallo} S, {Davinson} T, {Depalo} R, {Di Leva} A, {Elekes} Z, {Ferraro}
  F, {Fiore} E, {Formicola} A, {F{\"u}l{\"o}p} Z, {Gervino} G, {Guglielmetti}
  A, {Gustavino} C, {Gy{\"u}rky} G, {Imbriani} G, {Junker} M, {Lugaro} M,
  {Marigo} P, {Masha} E, {Menegazzo} R, {Mossa} V, {Pantaleo} F~R, {Paticchio}
  V, {Perrino} R, {Piatti} D, {Prati} P, {Schiavulli} L, {St{\"o}ckel} K,
  {Straniero} O, {Sz{\"u}cs} T, {Tak{\'a}cs} M~P, {Terrasi} F, {Vescovi} D,
  {Zavatarelli} S and {LUNA Collaboration} 2021 {\em Phys. Rev. Lett.\/} {\bf
  127} 152701 (\textit{Preprint} \eprint{2110.00303})

\bibitem{Bemmerer_et_al._2005}
Bemmerer D, Confortola F, Lemut A, Bonetti R, Broggini C, Corvisiero P,
  Costantini H, Cruz J, Formicola A, F{\"u}l{\"o}p Z, Gervino G, Guglielmetti
  A, Gustavino C, Gy{\"u}rky G, Imbriani G, Jesus A~P, Junker M, Limata B,
  Menegazzo R, Prati P, Roca V, Rogalla D, Rolfs C, Romano M, Rossi~Alvarez C,
  Sch{\"u}mann F, Somorjai E, Straniero O, Strieder F, Terrasi F, Trautvetter
  H~P and Vomiero A 2005 {\em EPJ A\/} {\bf 24} 313--319 ISSN 1434-601X

\bibitem{best-2016}
{Best} A, {G{\"o}rres} J, {Junker} M, {Kratz} K~L, {Laubenstein} M, {Long} A,
  {Nisi} S, {Smith} K and {Wiescher} M 2016 {\em Nucl. Instr. Meth. Phys. Res.
  A\/} {\bf 812} 1--6 (\textit{Preprint} \eprint{1509.00770})

\bibitem{Tilley_1995}
{Tilley} D~R, {Weller} H~R, {Cheves} C~M and {Chasteler} R~M 1995 {\em Nucl.
  Phys. A\/} {\bf 595} 1--170

\bibitem{Bruno-2016PhRvL}
{Bruno} C~G, {Scott} D~A, {Aliotta} M, {Formicola} A, {Best} A, {Boeltzig} A,
  {Bemmerer} D, {Broggini} C, {Caciolli} A, {Cavanna} F, {Ciani} G~F,
  {Corvisiero} P, {Davinson} T, {Depalo} R, {Di Leva} A, {Elekes} Z, {Ferraro}
  F, {F{\"u}l{\"o}p} Z, {Gervino} G, {Guglielmetti} A, {Gustavino} C,
  {Gy{\"u}rky} G, {Imbriani} G, {Junker} M, {Menegazzo} R, {Mossa} V,
  {Pantaleo} F~R, {Piatti} D, {Prati} P, {Somorjai} E, {Straniero} O,
  {Strieder} F, {Sz{\"u}cs} T, {Tak{\'a}cs} M~P, {Trezzi} D and {LUNA
  Collaboration} 2016 {\em Phys. Rev. Lett.\/} {\bf 117} 142502
  (\textit{Preprint} \eprint{1610.00483})

\bibitem{Lugaro-2017NatAs}
{Lugaro} M, {Karakas} A~I, {Bruno} C~G, {Aliotta} M, {Nittler} L~R, {Bemmerer}
  D, {Best} A, {Boeltzig} A, {Broggini} C, {Caciolli} A, {Cavanna} F, {Ciani}
  G~F, {Corvisiero} P, {Davinson} T, {Depalo} R, {di Leva} A, {Elekes} Z,
  {Ferraro} F, {Formicola} A, {F{\"u}l{\"o}p} Z, {Gervino} G, {Guglielmetti} A,
  {Gustavino} C, {Gy{\"u}rky} G, {Imbriani} G, {Junker} M, {Menegazzo} R,
  {Mossa} V, {Pantaleo} F~R, {Piatti} D, {Prati} P, {Scott} D~A, {Straniero} O,
  {Strieder} F, {Sz{\"u}cs} T, {Tak{\'a}cs} M~P and {Trezzi} D 2017 {\em Nature
  Astronomy\/} {\bf 1} 0027 (\textit{Preprint} \eprint{1703.00276})

\bibitem{Straniero-2017A&A}
{Straniero} O, {Bruno} C~G, {Aliotta} M, {Best} A, {Boeltzig} A, {Bemmerer} D,
  {Broggini} C, {Caciolli} A, {Cavanna} F, {Ciani} G~F, {Corvisiero} P,
  {Cristallo} S, {Davinson} T, {Depalo} R, {Di Leva} A, {Elekes} Z, {Ferraro}
  F, {Formicola} A, {F{\"u}l{\"o}p} Z, {Gervino} G, {Guglielmetti} A,
  {Gustavino} C, {Gy{\"u}rky} G, {Imbriani} G, {Junker} M, {Menegazzo} R,
  {Mossa} V, {Pantaleo} F~R, {Piatti} D, {Piersanti} L, {Prati} P, {Samorjai}
  E, {Strieder} F, {Sz{\"u}cs} T, {Tak{\'a}cs} M~P and {Trezzi} D 2017 {\em
  Astronomy \& Astrophysics\/} {\bf 598} A128 (\textit{Preprint}
  \eprint{1611.00632})

\bibitem{Mak-1980}
{Mak} H~B, {Ewan} G~T, {Evans} H~C, {MacArthur} J~D, {McLatchie} W and {Azuma}
  R~E 1980 {\em Nucl. Phys. A\/} {\bf 343} 79--90

\bibitem{Blackmon-1995}
{Blackmon} J~C, {Champagne} A~E, {Hofstee} M~A, {Smith} M~S, {Downing} R~G and
  {Lamaze} G~P 1995 {\em Phys. Rev. Lett.\/} {\bf 74} 2642--2645

\bibitem{Sergi-2010}
{Sergi} M~L, {Spitaleri} C, {La Cognata} M, {Coc} A, {Mukhamedzhanov} A,
  {Burjan} S~V, {Cherubini} S, {Crucill{\'a}} V, {Gulino} M, {Hammache} F,
  {Hons} Z, {Irgaziev} B, {Kiss} G~G, {Kroha} V, {Lamia} L, {Pizzone} R~G,
  {Puglia} S~M~R, {Rapisarda} G~G, {Romano} S, {de S{\'e}r{\'e}ville} N,
  {Somorjai} E, {Tudisco} S and {Tumino} A 2010 {\em Phys. Rev. C\/} {\bf 82}
  032801

\bibitem{Hannam-1999}
{Hannam} M~D and {Thompson} W~J 1999 {\em Nucl. Instr. Meth. Phys. Res. A\/}
  {\bf 431} 239--251

\bibitem{Caciolli2012}
{Caciolli} A, {Scott} D~A, {Di Leva} A, {Formicola} A, {Aliotta} M, {Anders} M,
  {Bellini} A, {Bemmerer} D, {Broggini} C, {Campeggio} M, {Corvisiero} P,
  {Depalo} R, {Elekes} Z, {F{\"u}l{\"o}p} Z, {Gervino} G, {Guglielmetti} A,
  {Gustavino} C, {Gy{\"u}rky} G, {Imbriani} G, {Junker} M, {Marta} M,
  {Menegazzo} R, {Napolitani} E, {Prati} P, {Rigato} V, {Roca} V, {Rolfs} C,
  {Rossi Alvarez} C, {Somorjai} E, {Salvo} C, {Straniero} O, {Strieder} F,
  {Sz{\"u}cs} T, {Terrasi} F, {Trautvetter} H~P and {Trezzi} D 2012 {\em EPJ
  A\/} {\bf 48} 144 (\textit{Preprint} \eprint{1210.0327})

\bibitem{DiLeva2014}
Di~Leva A, Scott D~A, Caciolli A, Formicola A, Strieder F, Aliotta M, Anders M,
  Bemmerer D, Broggini C, Corvisiero P, Elekes Z, F\"ul\"op Z, Gervino G,
  Guglielmetti A, Gustavino C, Gy\"urky G, Imbriani G, Jos\'e J, Junker M,
  Laubenstein M, Menegazzo R, Napolitani E, Prati P, Rigato V, Roca V, Somorjai
  E, Salvo C, Straniero O, Sz\"ucs T, Terrasi F and Trezzi D (LUNA
  Collaboration) 2014 {\em Phys. Rev. C\/} {\bf 89}(1) 015803

\bibitem{Asakawa-2020JVSTB}
{Asakawa} T, {Nagano} D, {Miyazawa} H and {Clark} I 2020 {\em J. Vac. Sci.
  Technol. B\/} {\bf 38} 034008

\bibitem{gyurky2019}
{Gy{\"u}rky} G, {F{\"u}l{\"o}p} Z, {K{\"a}ppeler} F, {Kiss} G~G and {Wallner} A
  2019 {\em EPJ A\/} {\bf 55} 41 (\textit{Preprint} \eprint{1903.03339})

\bibitem{bemmerer_activation_2006}
Scott D~A, Caciolli A, Di~Leva A, Formicola A, Aliotta M, Anders M, Bemmerer D,
  Broggini C, Campeggio M, Corvisiero P, Elekes Z, F\"ul\"op Z, Gervino G,
  Guglielmetti A, Gustavino C, Gy\"urky G, Imbriani G, Junker M, Laubenstein M,
  Menegazzo R, Marta M, Napolitani E, Prati P, Rigato V, Roca V, Somorjai E,
  Salvo C, Straniero O, Strieder F, Sz\"ucs T, Terrasi F and Trezzi D (LUNA
  Collaboration) 2012 {\em Phys. Rev. Lett.\/} {\bf 109}(20) 202501

\bibitem{Scott2012}
Scott D~A, Caciolli A, Di~Leva A, Formicola A, Aliotta M, Anders M, Bemmerer D,
  Broggini C, Campeggio M, Corvisiero P, Elekes Z, F\"ul\"op Z, Gervino G,
  Guglielmetti A, Gustavino C, Gy\"urky G, Imbriani G, Junker M, Laubenstein M,
  Menegazzo R, Marta M, Napolitani E, Prati P, Rigato V, Roca V, Somorjai E,
  Salvo C, Straniero O, Strieder F, Sz\"ucs T, Terrasi F and Trezzi D (LUNA
  Collaboration) 2012 {\em Phys. Rev. Lett.\/} {\bf 109}(20) 202501

\bibitem{gyurky_activation_2019}
Gy\"urky G, Hal\'asz Z, Kiss G~G, Sz\"ucs T, Cs\'{\i}k A, T\"or\"ok Z,
  Husz\'ank R, Kohan M~G, Wagner L and F\"ul\"op Z 2019 {\em Phys. Rev. C\/}
  {\bf 100}(1) 015805

\bibitem{gyurky_activation_2022}
Gyürky G, Halász Z, Kiss G~G, Szücs T and Fülöp Z 2022 {\em Phys. Rev.
  C\/} {\bf 105} L022801 ISSN 2469-9985, 2469-9993

\bibitem{SRIM2003}
{Ziegler} J~F 2004 {\em Nucl. Instr. Meth. Phys. Res. B\/} {\bf 219} 1027--1036

\bibitem{di_leva_underground_2014}
Di~Leva A, Scott D~A, Caciolli A, Formicola A, Strieder F, Aliotta M, Anders M,
  Bemmerer D, Broggini C, Corvisiero P, Elekes Z, Fülöp Z, Gervino G,
  Guglielmetti A, Gustavino C, Gyürky G, Imbriani G, José J, Junker M,
  Laubenstein M, Menegazzo R, Napolitani E, Prati P, Rigato V, Roca V, Somorjai
  E, Salvo C, Straniero O, Szücs T, Terrasi F and Trezzi D (LUNA) 2014 {\em
  Phys. Rev. C\/} {\bf 89} ISSN 0556-2813, 1089-490X

\end{thebibliography}


\end{document}